\newcommand{\healpix}{\tt\string HEALPix}

\newcommand{\sqd}{deg$^{2}$}

\documentclass[twocolumn]{aastex631}

\usepackage{hyperref}
\usepackage{natbib}
\usepackage{url}
\providecommand{\qcr}[1]{\texttt{#1}}
\providecommand{\software}[1]{#1}

\begin{document}
\renewcommand{\arraystretch}{0.9}

\title{Las Cumbres Observatory Gravitational-Wave Follow-up in the Third and Fourth Observing Runs: Strengths and Weaknesses of a Rapid Response Galaxy Targeted Strategy}

\author[0000-0002-0675-0887]{Ido Keinan}
\affiliation{The School of Physics and Astronomy, Tel Aviv University, Tel Aviv 69978, Israel}

\author[0000-0001-7090-4898]{Iair Arcavi}
\affiliation{The School of Physics and Astronomy, Tel Aviv University, Tel Aviv 69978, Israel}

\author[0000-0003-4253-656X]{D. Andrew Howell}
\affiliation{Las Cumbres Observatory, 6740 Cortona Drive, Suite 102, Goleta, CA 93117-5575, USA}
\affiliation{Department of Physics, University of California, Santa Barbara, CA 93106-9530, USA}

\author[0000-0001-5807-7893]{Curtis McCully}
\affiliation{Las Cumbres Observatory, 6740 Cortona Drive, Suite 102, Goleta, CA 93117-5575, USA}

\author[0000-0002-7472-1279]{Craig Pellegrino}
\affiliation{Goddard Space Flight Center, 8800 Greenbelt Road, Greenbelt, MD 20771, USA}

\author{Ayelet Hasson}
\affiliation{Department of Physics of Complex Systems, Weizmann Institute of Science, Rehovot 76100, Israel}

\author[0000-0002-1895-6639]{Moira Andrews}
\affiliation{Las Cumbres Observatory, 6740 Cortona Drive, Suite 102, Goleta, CA 93117-5575, USA}
\affiliation{Department of Physics, University of California, Santa Barbara, CA 93106-9530, USA}

\author[0000-0003-0035-6659]{Jamison Burke}
\affiliation{Shady Side Academy, 423 Fox Chapel Road, Pittsburgh, PA 15238, USA}

\author[0000-0002-1125-9187]{Daichi Hiramatsu}
\affiliation{Department of Astronomy, University of Florida 211 Bryant Space Science Center, Gainesville, FL 32611-2055 USA}
\affiliation{Center for Astrophysics | Harvard \& Smithsonian, 60 Garden Street, Cambridge, MA 02138-1516, USA}
\affiliation{The NSF AI Institute for Artificial Intelligence and Fundamental Interactions, USA}

\author[0000-0003-3340-4784]{Jennifer Barnes}
\affiliation{Kavli Institute for Theoretical Physics, Kohn Hall, University of California, Santa Barbara, CA 93106, USA}

\author[0000-0001-6711-8140]{Sukanya Chakrabarti}
\affiliation{Department of Physics and Astronomy, University of Alabama in Huntsville, 301 North Sparkman Drive, Huntsville, AL 35816, USA}

\author[0000-0003-4914-5625]{Joseph R. Farah}
\affiliation{Las Cumbres Observatory, 6740 Cortona Drive, Suite 102, Goleta, CA 93117-5575, USA}
\affiliation{Department of Physics, University of California, Santa Barbara, CA 93106-9530, USA}

\author[0000-0002-4488-726X]{Paul J. Groot}
\affiliation{Department of Astrophysics/IMAPP, Radboud University, PO Box 9010, 6500 GL Nijmegen, The Netherlands}
\affiliation{Department of Astronomy, University of Cape Town, Private Bag X3, Rondebosch, 7701, South Africa}
\affiliation{South African Astronomical Observatory, P.O. Box 9, Observatory, 7935, South Africa}
\affiliation{The Inter-University Institute for Data Intensive Astronomy, University of Cape Town, Private Bag X3, Rondebosch, 7701, South
Africa}

\author[0000-0002-0430-7793]{Na'ama Hallakoun}
\affiliation{Department of Particle Physics and Astrophysics, Weizmann Institute of Science, Rehovot 7610001}

\author[0000-0002-0175-5064]{Daniel Holz}
\affiliation{Department of Physics, Department of Astronomy \& Astrophysics, The University of Chicago, 5640 South Ellis Avenue, Chicago, IL 60637, USA
}

\author[0000-0001-8738-6011]{Saurabh W. Jha}
\affiliation{Department of Physics and Astronomy, Rutgers, State University of New Jersey, 136 Frelinghuysen Road, Piscataway, NJ 08854, USA}

\author[0000-0002-5981-1022]{Daniel Kasen}
\affiliation{Department of Astronomy and Theoretical Astrophysics Center, University of California, Berkeley, CA 94720, USA}

\author[0000-0003-1731-0497]{Chris Lidman}
\affiliation{The Research School of Astronomy and Astrophysics, Australian National University, Stromlo, ACT, Australia}

\author[0000-0001-9589-3793]{Michael J. Lundquist}
\affiliation{W.M. Keck Observatory, 65-1120 Mamalahoa Highway, Kamuela, HI 96743, USA}

\author[0000-0002-6579-0483]{Dan Maoz}
\affiliation{The School of Physics and Astronomy, Tel Aviv University, Tel Aviv 69978, Israel}

\author[0000-0002-4670-7509]{Brian D. Metzger}
\affiliation{Center for Computational Astrophysics, Flatiron Institute, 162 5th Avenue, New York, NY 10010, USA}
\affiliation{Department of Physics and Columbia Astrophysics Laboratory, Columbia University, New York, NY 10027, USA}

\author[0000-0002-4534-7089]{Ehud Nakar}
\affiliation{The School of Physics and Astronomy, Tel Aviv University, Tel Aviv 69978, Israel}

\author[0000-0001-9570-0584]{Megan Newsome}
\affiliation{University of Texas at Austin, 1 University Station C1400, Austin, TX 78712-0259, USA}

\author[0000-0003-3656-5268]{Yuan Qi Ni}
\affiliation{Kavli Institute for Theoretical Physics, University of California, Santa Barbara, 552 University Road, Goleta, CA 93106-4030, USA}
\affiliation{Las Cumbres Observatory, 6740 Cortona Drive, Suite 102, Goleta, CA 93117-5575, USA}

\author[0000-0002-1850-4587]{Alexander H. Nitz}
\affiliation{Department of Physics, Syracuse University, Syracuse, NY 13244, USA}

\author[0000-0003-0209-9246]{Estefania Padilla Gonzalez}
\affiliation{Space Telescope Science Institute, Baltimore, MD 21218, USA}

\author[0000-0002-7964-5420]{Tsvi Piran}
\affiliation{Racah Institute for Physics, The Hebrew University, Jerusalem 91904, Israel}

\author[0000-0003-1470-7173]{Dovi Poznanski}
\affiliation{The School of Physics and Astronomy, Tel Aviv University, Tel Aviv 69978, Israel}
\affiliation{Cahill Center for Astrophysics, California Institute of Technology, Pasadena CA 91125, USA}
\affiliation{Kavli Institute for Particle Astrophysics \& Cosmology, 452 Lomita Mall, Stanford University, Stanford, CA 94305, USA}
\affiliation{Department of Physics, Stanford University, 382 Via Pueblo Mall, Stanford, CA 94305, USA}

\author[0000-0003-1724-2885]{Ryan Ridden-Harper}
\affiliation{School of Physical and Chemical Sciences — Te Kura Matu, University of Canterbury, Private Bag 4800, Christchurch 8140, New Zealand}

\author[0000-0003-4102-380X]{David J. Sand}
\affiliation{Steward Observatory, University of Arizona, 933 North Cherry Avenue, Tucson, AZ 85721-0065, USA}

\author[0000-0001-6589-1287]{Brian P. Schmidt}
\affiliation{Research School of Astronomy and Astrophysics, Australian National University, Canberra, ACT 2611, Australia}
\affiliation{ARC Centre of Excellence for All-sky Astrophysics (CAASTRO), Australian National University, Canberra, ACT 2611, Australia}

\author[0000-0003-0794-5982]{Giacomo Terreran}
\affiliation{Adler Planetarium, 1300 S DuSable Lake Shore Dr, Chicago, IL 60605, USA}

\author[0000-0002-4283-5159]{Brad E. Tucker}
\affiliation{The Research School of Astronomy and Astrophysics, Australian National
University, ACT 2601, Australia}

\author[0000-0001-8818-0795]{Stefano Valenti}
\affiliation{Department of Physics, University of California, 1 Shields Avenue, Davis, CA 95616-5270, USA}

\author[0000-0003-1349-6538]{J. Craig Wheeler}
\affiliation{Department of Astronomy, University of Texas at Austin, 2515 Speedway Stop C1400, Austin, TX 78712-1205, USA}

\author[0000-0003-2732-4956]{Samuel Wyatt}
\affiliation{Dr. Robert H. Goddard Space Flight Center NASA/GSFC ICESat Road, Greenbelt Road Greenbelt, Maryland 20771-0001}
\correspondingauthor{Ido Keinan}
\email{idokeinan1@mail.tau.ac.il}

\author[0009-0006-7296-728X]{Kathryn Wynn}
\affiliation{Las Cumbres Observatory, 6740 Cortona Drive, Suite 102, Goleta, CA 93117-5575, USA}
\affiliation{Department of Physics, University of California, Santa Barbara, CA 93106-9530, USA}

\begin{abstract}
\label{abstract}
We present a summary of gravitational-wave (GW) follow-up using the Las Cumbres Observatory global network of telescopes during the third (O3) and fourth (O4) observing runs of the GW detectors. As in O2, we implemented the \cite{Gehrels_2016_galaxy_strategy} galaxy-targeted strategy. Here we test its efficacy in O3 and O4 and analyze the Las Cumbres Observatory response time and depth for nine GW alerts that showed a possibility of having an electromagnetic counterpart (GW190425, GW190426\_152155, S190510g, GW190728\_064510, GW190814, S190822c, GW191216\_213338, S240422ed and S250206dm). We find that Las Cumbres Observatory is able to begin observations in response to GW alerts within minutes of the alert, with the observations being deep enough to detect possible GW170817-like kilonovae out to a median distance of 250 Mpc. In this sense a global rapid-response network of telescopes like Las Cumbres is an excellent GW follow-up facility. However, the galaxy-targeted follow-up strategy was much less efficient in O3 and O4 than originally predicted, given the larger than assumed GW localizations. We conclude that coordination between various facilities to include both wide-field and rapid-response capabilities is required to achieve efficient and comprehensive follow-up of GW events.
\end{abstract}


\section{Introduction}
\label{sec: intro}
The second observing run (O2) of the Advanced Laser Interferometer Gravitational-wave Observatory \citep[LIGO;][]{LIGO_Scientific_Collaboration_2015, Davis_2021_ligo_char_o2_o3} and Advanced Virgo \citep{Acernese_2015_adv_virgo}, known then as the LIGO-Virgo science Collaboration (LVC), took place from November 30, 2016, to August 25, 2017, and resulted in the first detection of a binary neutron star (BNS) merger \citep{gw170817,gwtc_1}. This gravitational-wave (GW) event, GW170817, was followed by the (so far only) confirmed discovery of electromagnetic (EM) counterparts, which include the short Gamma Ray Burst \citep[GRB;][]{Goldstein_2017, Savchenko_2017} GRB170817A detected $\sim2$ seconds after the merger; X-ray and radio afterglows \citep{Hallinan_2017, troja_gw170817_obs, Alexander_2018, Margutti_2018, pozanenko_gw170817_obs} detected days to weeks later; and a kilonova, identified approximately 11 hours after the 
merger (\citealt{andreoni_gw170817_obs}; \citealt{arcavi_gw170817_obs}; \citealt{coulter_gw170817_obs, cowperthwaite_gw170817_obs}; \citealt{diaz_gw170817_obs}; \citealt{drout_gw170817_obs, evans_gw170817_obs, hu_gw170817_obs, kasliwal_gw170817_obs}; \citealt{lipunov_gw170817_obs, lco_spectra_gw170817_mccully}; \citealt{pian_gw170817_obs, shapee_gw170817_obs, smartt_gw170817_obs}; \citealt{decam_gw170817_soares_santos, tanvir_gw170817_obs, utsumi_gw170817_obs, dlt40_gw170817_valenti,salt_gw170817_buckley}; \citealt{pozanenko_gw170817_obs}). This discovery provided a wealth of information, including the association of BNS mergers with short GRBs \citep{Abbott2017_multi_messenger, Goldstein_2017, Savchenko_2017}, evidence for an off-axis structured jet geometry from the afterglow evolution \citep{Margutti_2018, Mooley_2018, Ghirlanda_2019}, and confirmation that BNS mergers are a site of rapid ($r$-)process nucleosynthesis \citep{drout_gw170817_obs, kasen2017, pian_gw170817_obs}. However, it remains unclear whether GW170817 is representative of the broader BNS merger population. Models predict a wide diversity of possible transients driven by variation in merger component masses, ejecta mass and composition, viewing angle, and uncertain physics such as the neutron star equation of state  \citep[e.g.][]{kasen2017, bulla_2019_model_possis, nicholl_2021_models}. The 11 hour delay in the discovery of the GW170817 kilonova made it challenging to discriminate among kilonova emission models \cite[e.g.][]{Arcavi_early_hours}, demonstrating that even faster follow-up is required. For reviews of GW170817 see \cite{metzger2017}, \cite{nakar2020}, and \cite{Margutti2021}. 

The third observing run (O3), operated by the LIGO, VIRGO, and the Kamioka Gravitational Wave detector \citep[KAGRA;][]{abe2022_kagra_interferometer} collaboration (LVK), ran from April 1, 2019 to October 1, 2019 (O3a), and from November 1, 2019 to March 27, 2020 (O3b). O3 included four detectors: the two advanced LIGO interferometers, in Livingston, Louisiana, USA (L1) and in Hanford, Washington, USA (H1), the advanced Virgo detector in Cascina, Italy (V1), and the KAGRA detector in Hida, Japan (K1), which began operations towards the end of O3b. The median BNS range \citep{gw_distances_chen_2021} of each instrument during O3a (O3b) was 135 (133) Mpc, 108 (155) Mpc, and 45 (51) Mpc, for L1, H1, and V1, respectively \citep{gwtc_2, abbott2023_gwtc3}. K1 had a BNS range of 0.7 Mpc at the end of O3b\footnote{\url{https://observing.docs.ligo.org/plan/}}. During O3, alerts were issued for 80 significant events, defined as candidates with a False Alarm Rate (FAR) below one alert per seven months, which is the threshold adopted by the LVK for compact binary coalescence (CBC) searches\footnote{\url{https://emfollow.docs.ligo.org/userguide/analysis/index.html\#alert-threshold}}. Early-warning alerts, issued prior to merger, were classified as significant if they had a FAR below one alert per four months\footnote{\url{https://emfollow.docs.ligo.org/userguide/early_warning.html}}. Alerts with higher FAR values were classified as low-significance events. Of the 80 significant alerts, 24 were later retracted following improved analyses indicating a non-astrophysical origin, leaving a total of 54 confirmed significant events \citep{gwtc_2, abbott2023_gwtc3, abbott2022_gwtc2_1}.

The fourth observing run (O4), operated by the LVK, ran from May 24, 2023 to January 16, 2024 (O4a), April 10, 2024 to January 28, 2025 (O4b), and January 28, 2025 to November 18, 2025 (O4c). The L1 and H1 detectors operated during O4a, O4b, and O4c, while V1 operated only during O4b and O4c. K1 operated at the beginning of O4a and during O4c \citep{gwtc_4a}. The BNS ranges of each instrument during O4 were 150-160 Mpc for L1 and H1, 50-60 Mpc for V1, and 1-10 Mpc for K1. During O4, alerts were issued for 283 significant events, of which 29 were retracted, leaving a total of 254 confirmed significant events, and 5136 alerts for low significance events.

We used the Las Cumbres Observatory global network of telescopes \citep[hereafter, Las Cumbres;][]{lco_capabilities} as part of a Key Project (PIs: Arcavi and Howell) to follow seven events during O3, of which five initially showed a non-negligible probability for an EM counterpart and two were of lower confidence\footnote{For simplicity, we collectively refer to all of these as ``events'', even if they were later determined not to be astrophysical in origin.}. Of these, only one \citep[GW190425;][]{gw190425_abbott2020} remained as an event with a probable EM counterpart after final LVK analysis \citep{gwtc_2, abbott2023_gwtc3, abbott2022_gwtc2_1}, with the rest having either low or no probability of an EM counterpart (we keep these events in our analysis since they still serve as a test of the follow-up performance). During O4, follow-up observations were obtained for 2 significant events that had a non-negligible probability for an EM counterpart (final analysis of these events by the LVK is not yet available).

Here we analyze the follow-up performance of the Las Cumbres search for EM counterparts to the nine GW events followed by us during O3 and O4. We briefly describe the relevant Las Cumbres Observatory facilities in Section \ref{sec:lco} and the GW follow-up strategy in Section \ref{sec:followup_strategy}. Section \ref{sec:observations} details the events followed and the follow-up observations. The follow-up performance analysis is described in Section \ref{sec:analysis}, and its results are presented in Section \ref{sec:result_and_discussion}. Our summary and conclusions are presented in Section \ref{sec:summary}.

\section{Las Cumbres Observatory}
\label{sec:lco}
During O3 and O4, 19 optical telescopes (two 2\,m, nine 1\,m, and eight 0.4\,m in diameter) at five of the Las Cumbres sites took part in GW follow-up (Table \ref{tab:lco_observatories}) as part of our Key Project. All Las Cumbres telescopes are automated and the entire network shares a dynamic schedule. The observatory capabilities are described in detail in \cite{lco_capabilities}.

Each telescope class uses a different type of imager with a different field of view (FOV) and pixel scale (Table \ref{tab:imagers_proparties}). All imagers used here are paired with Sloan Digital Sky Survey (SDSS) $ugriz$ filters, Johnson $BVRI$ filters, and a wide $w$ filter covering the $gri$ bands. 

\begin{deluxetable}{llll}
\label{tab:lco_observatories}
\tablecaption{Las Cumbres Telescopes relevant for this work, ordered from east to west.}
\tablehead{
\colhead{Observatory}\hspace{0.025\textwidth} &
\colhead{Location}\hspace{0.025\textwidth} &
\colhead{Code}\hspace{0.025\textwidth} &
\colhead{Telescopes}
}
\startdata
Siding Spring & Australia & COJ & 2 m (\ensuremath{\times}1)\\
& & & 1\,m (\ensuremath{\times}2)\\
& & & 0.4\,m (\ensuremath{\times}2)\\
SAAO & South Africa & CPT & 1\,m (\ensuremath{\times}3)\\
& & & 0.4\,m (\ensuremath{\times}1)\\
CTIO & Chile & LSC & 1\,m (\ensuremath{\times}3)\\
& & & 0.4\,m (\ensuremath{\times}2)\\
McDonald & Texas, USA & ELP & 1\,m (\ensuremath{\times}1) \\
& & & 0.4\,m (\ensuremath{\times}1) \\
Haleakala & Hawaii, USA & OGG & 2\,m (\ensuremath{\times}1) \\ 
& & & 0.4\,m (\ensuremath{\times}2) \\
\enddata
\tablecomments{The number of telescopes of each class in each site is indicated in parenthesis in the right column.}
\end{deluxetable}

\begin{deluxetable}{lcccc}
\label{tab:imagers_proparties}
\tablecaption{Las Cumbres imagers relevant for this work.}
\tablehead{
\colhead{Class} &
\colhead{Imager} &
\colhead{FOV} &
\colhead{Pixel Scale} & 
\colhead{Run} \\ & & & (Binning) &
}
\startdata
2\,m & Spectral & $10'\times10'$ & 0.3 ($2\times2$) & O3 \\
& MuSCAT$^{1}$ &  $9.1'\times9.1'$ & 0.27 ($1\times1$) & O4 \\
1\,m & Sinistro & $26'\times26'$ & 0.389 ($1\times1$)& O3, O4 \\
0.4\,m & SBIG6303 & $29'\times19'$ & 0.571 ($1\times1$) & O3 \\
& QHY600 & $1.9^{\circ}\times1.2^{\circ}$ & 0.74 ($1\times1$) & O4
\enddata
\tablenotetext{}{$^{1}$Multi-color Simultaneous Camera for studying Atmospheres of Transiting exoplanets \citep{Narita2020}.}
\tablecomments{The pixel scale is given in arcseconds per pixel after binning.}
\end{deluxetable}

\section{The Follow-up Strategy}
\label{sec:followup_strategy}
During O3, GW events determined to be significant by real-time analysis were distributed as alerts through the Gamma-ray Coordinates Network\footnote{\url{https://gcn.nasa.gov}} (GCN; now the General Coordinates Network) notices. These notices include the event type\footnote{Binary black hole (BBH) merger, neutron star-black hole (NSBH) merger, BNS merger, a merger that contains a ``mass gap'' object with a mass of $3M_{\odot}\lesssim M\lesssim 5M_{\odot}$, or an event of ``terrestrial'' origin, i.e. noise from Earth.} probabilities, the probability of the event being of astrophysical origin $p_{astro}$, the FAR of the signal, and a full Hierarchical Equal
Area isoLatitude Pixelation\footnote{\url{https://HEALPix.sourceforge.io/}} \citep[HEALPix;][]{gorski_2005_heapix} sky map of location probabilities and distance estimates. The HEALPix map divides the sky into equal-area pixels, with each pixel containing a probability for the GW source to be at that location, along with a mean distance estimate for the GW source if it were at that position, and its uncertainty. The total localization area and volume can be calculated, for example, with the {\tt{healpy}}\footnote{\url{https://healpy.readthedocs.io/en/latest/}} Python package. 

The first alerts to be released for each event were ``preliminary'' (not yet human vetted), followed by ``initial'' (issued after human vetting) alerts. Most ``initial'' alerts were then followed by one or more ``update'' alerts (based on more detailed analysis of the GW signal), which updated one or more of the alert parameters, or a ``retraction'' alert (if the event was no longer deemed real). 

The VOEvent protocol \citep{voevent_seaman2011} was used to automatically receive the alerts, store them in a database, and trigger galaxy selection. An email notification was sent to all team members for all significant alerts, and an additional text message was sent for events having a probability greater than 10\% of containing a neutron star or mass-gap object.
Target of Opportunity (ToO) observations were then manually triggered for events having $p_{astro}>50\%$, having a probability greater than 10\% of containing a neutron star or mass-gap object, observable for at least two hours per night, and for which the Las Cumbres network was not suffering from technical outages at the time. Continued observations were re-evaluated following each update alert. Observations were stopped after a few days, depending on the event, as detailed in Section \ref{subsec:response_time}.

Due to the FOV of the Las Cumbres imagers (Table \ref{tab:imagers_proparties}), galaxy-targeted observations, where telescope pointings are centered on specific galaxies in the GW localization region, are more efficient than trying to tile the entire GW localization region. Here, as in O2 \citep{Arcavi_2017_o2_strategy}, we follow the galaxy-targeted approach presented in \cite{Gehrels_2016_galaxy_strategy}. In practice, for each alert we calculated the localization volume by integrating over the 3$\sigma$ distance estimate for every {\healpix} pixel within the 99\% localization probability. Galaxies from the Galaxy List for the Advanced Detector Era\footnote{\url{https://glade.elte.hu/}} \citep[GLADE;][]{glade_catalog} catalog Version 2.3 inside this volume were prioritized and assigned a ``score'' for the likelihood of detecting the counterpart there, based on their location, mass, distance, and our limiting detectable luminosity, as described in \cite{Arcavi_2017_o2_strategy}. \cite{Arcavi_2017_o2_strategy} show that Version 2 of GLADE is complete out to $\sim$300 Mpc for galaxies with $L_B/L^{*}_B$ greater than the Schechter function \citep{schechter_function} median value, with $L_B$ the $B$-band galaxy luminosity and $L^{*}_B$ the luminosity equivalent to an absolute magnitude of $M^{*}_{B}=-20.7$ (see \citealt{Arcavi_2017_o2_strategy} for more details). In O3 we removed galaxies fainter than this median value from the prioritization, while in O4, this luminosity restriction was relaxed, to allow for a larger pool of galaxies. 

The GLADE catalog is based on a compilation of several other catalogs\footnote{The 2 Micron All-Sky Survey Extended Source Catalog \citep[2MASS XSC;][]{skrutskie_2006_2mass}, the Gravitational Wave Galaxy Catalog \citep[GWGC;][]{white_2011_gwgc}, the 2MASS Photometric Redshift Catalog \citep[2MPZ;][]{bilicki_2014_2mpz}, the combination of Hypercat and the Lyon-Meudon Extragalactic Database \citep[HyperLEDA;][]{makarov_2014_hyperleda}, and the Sloan Digital Sky Survey quasar catalogue from the 12th data release \citep[SDSS-DR12Q;][]{paris_2017_sdssdr12q}}. As a result, luminosity distance estimates are derived using heterogeneous methods, and associated uncertainties are not provided. Our galaxy selection algorithm thus adopts the luminosity distances as they are in the catalog without applying additional assumptions.

Las Cumbres observation triggers were then automatically prepared based on the galaxy prioritization, and were sent to the Las Cumbres real-time scheduler following human verification (see \citealt{Arcavi_2017_o2_strategy} for more details). Each galaxy was observed in multiple bands (see Section \ref{sec:observations}), either once or multiple times, depending on the observing constraints and localization size. Images taken were processed by the Beautiful Algorithms to Normalize Zillions of Astronomical Images (BANZAI) pipeline\footnote{\url{https://lco.global/documentation/data/BANZAIpipeline/}} \citep{banzai_mccully_2018} and then by the {\tt{LCOGTSNpipe}}\footnote{\url{https://github.com/LCOGT/lcogtsnpipe}} pipeline \citep{valenti2016_lcogtsnpipe} which also subtracts template images from the Sloan Digital Sky Survey (SDSS) Data Release 15 \citep{sdss_dr15_aguado_2019} when available, and calculates a global limiting magnitude per image. This estimate does not account for possible residual flux from the host galaxy due to imperfect subtractions nor for the higher background in bright host-galaxy regions. The true limit for transients located in bright regions of their host galaxy could therefore be shallower.

After processing, image triplet (new, reference and subtraction when available) cutouts around each target galaxy were presented on a webpage for manual inspection for possible new sources. When no references were available, only the new image was presented. In this case, a counterpart embedded in a bright region of the host galaxy might not be detected in real time.

\section{Alerts and Observations}
\label{sec:observations}
Each GW event identified by the LVK receives a GW Candidate Event Database (GraceDB)\footnote{\url{https://gracedb.ligo.org}} superevent ID (denoted with a leading `S'), and, if the event later surpasses $p_{astro}>50\%$ following final analysis, a GW Transient Catalog \citep[GWTC;][]{gwtc_2, abbott2023_gwtc3, abbott2022_gwtc2_1} ID (denoted with a leading `GW'). We use the GWTC ID if available. Otherwise, we use the GraceDB ID for the events presented in this paper. The GWTC/GraceDB IDs of the seven events followed by Las Cumbres During O3 are GW190425, GW190426\_152155, S190510g, GW190728\_064510, GW190814 \citep{gw190814_abbott2020}, S190822c and GW191216\_213338.
Of these events, one (S190822c) was retracted, and one (S190510g) was not retracted but does not appear in the GWTC due to having its $p_{astro}$ later re-evaluated to $<50\%$. The two events followed by Las Cumbres during O4 were S240422ed and S250206dm, detected during O4b and O4c, respectively. Events from these sub-runs have yet to be published in official GW catalogs. S240422ed had its $p_{astro}$ re-evaluated to $<50\%$ a few weeks after its detection \citep{gcn36812_s240422ed_lowsig}. In Table \ref{tab:events_summary} we provide a summary of all the alerts followed. We retrieve the alert information for GWTC events from the GW Open Science Center \citep[GWOSC;][]{gwosc_abbott_2023} website\footnote{\url{https://gwosc.org/eventapi/html/GWTC/}} and for non-GWTC events from GraceDB.

Las Cumbres observations of prioritized galaxies during O3 (O4) were taken in the $g$, $r$ and/or $i$ ($g$, $r$, $i$ and/or $z$) filters, with exposure times of 240 and 300 seconds (depending on the galaxy distance), and a 24-hour cadence. The choice of filters was motivated by the preference to obtain early multi-band color information during the critical initial hours \citep{Arcavi_early_hours}, even at the expense of maximizing sky coverage. This strategy enables the characterization of transient candidates at early times, even in cases where their identification is achieved in retrospect following discovery by more sensitive facilities. The adopted cadence represents a compromise between covering a sufficient number of prioritized galaxies and revisiting previously observed targets on the typical time scales for significant changes in emission properties, according to models. 

\begin{figure*} 
\centering
\vspace{-2cm}
\includegraphics[width=1.00\textwidth]{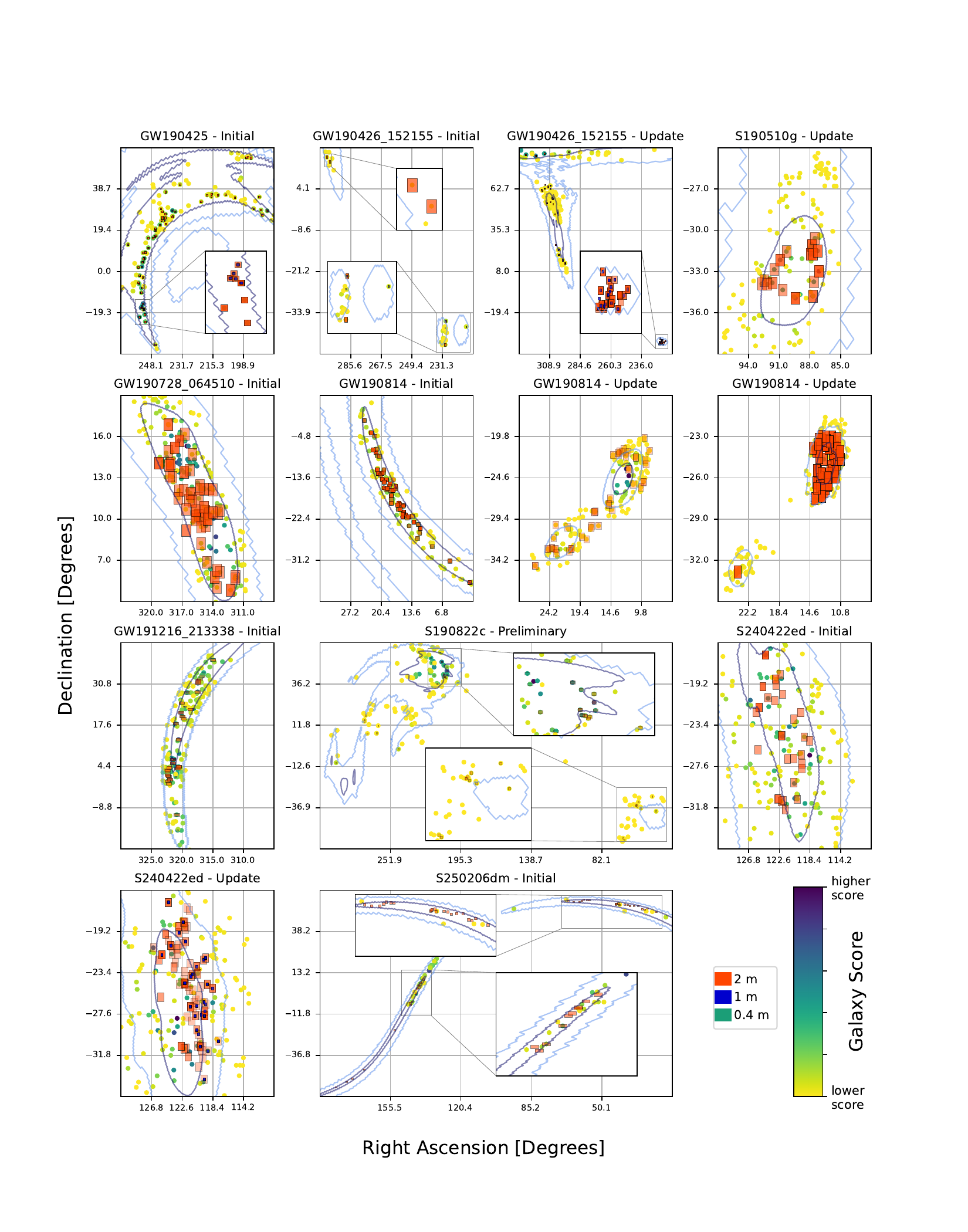}
\caption{Sky maps of events followed up by Las Cumbres. Insets show zoomed-in regions. Filled markers denote positions of galaxies colored by their probability score. Observed footprints of the 2\,m, 1\,m, and 0.4\,m telescopes are shown in red, blue, and green squares, respectively, denoting the true footprint size of their respective imagers. Light blue and purple contours are for the 90\% and 50\% localization regions, respectively. We show all the alerts for which observations were triggered.}
\label{fig:all_sky_plots}
\end{figure*}

A total of 1851 (525) images of 427 (118) galaxies were taken for the seven O3 (two O4) events. Of these, we removed 173 (162) images with exposures shorter than 100 seconds (which were due to technical or weather interrupts) and those for which the pipeline was not able to calculate a limiting magnitude (indicating a problem with the image), leaving 1678 (363) images\footnote{Images without a computed limiting magnitude could still reveal a transient if present. However, for consistency we restrict our analysis to images with measured limiting magnitudes.} of 419 (94) galaxies. We present the list of observations in Table \ref{tab:observations_semilist}, and in Figure \ref{fig:all_sky_plots} we present the footprints of these observations. All of our pointings also appear on the Treasure Map\footnote{\url{http://www.treasuremap.space}} \cite{wyatt2020_treasuremap}. Some of our follow-up was reported in real time through GCN circulars \citep{24488_s190510g_1, 24307_s190425z_5, 24249_s190426c_1, 24206_s190425z_2, 24194_s190425z_1, 25422_s190814bv_2, 25359_s190814bv_1, 24207_s190425z_3, 24225_s190425z_4, 36480_s240422ed_1}. 

During these campaigns, we also followed up two GW190425 counterpart candidates reported by \cite{2019GCN.24191_kasliwal_candidates} and identified an additional candidate during the follow-up of GW190426\_152155 \citep{24249_s190426c_1}. Subsequent analysis and community follow-up determined that all three candidates were unrelated to their respective GW events \citep{2019GCN.24251_unrelated_transient, 2019GCN.24205_buckley, 2019GCN.24208_izzo, 2019GCN.24211_nicholl, 2019GCN.24200_pavana, 2019GCN.24204_perley, 2019GCN.24209_wiersema}. No additional counterpart candidates  were identified in our observations. Our observations of GW190425 are also published in \cite{coulter2024_grav_col} and \cite{keinan2025}, and of GW190814 in \cite{kilpatrick2021_grav_col}.

\startlongtable
\begin{deluxetable*}{lcccccccc}
\label{tab:events_summary}
\centerwidetable
\tablecaption{Summary of alerts and Las Cumbres observations obtained per event followed here.}
\tablehead{
\colhead{Occasion} & \colhead{Time} & \colhead{90\% (50\%)} & \colhead{Distance} & $p_{astro}$ & \colhead{Type} & \colhead{FAR} & \colhead{No. of} & \colhead{No. of}\\
\colhead{} & \colhead{[UT]} & \colhead{Area [\sqd]} & \colhead{[Mpc]} & \colhead{} & \colhead{} & \colhead{[yr$^{-1}$]} & \colhead{Images} & \colhead{Galaxies}
}
\startdata
\hline
\hline
\multicolumn{9}{c}{Observation Run 3} \\
\hline
\hline
\multicolumn{9}{c}{GW190425} \\
\hline
GW Detection$^{1}$ & 2019-04-25 08:18:05 & & & & & & & \\
Initial Alert$^{2}$ & 2019-04-25 09:00:51 & 10183 (2806) & $155\pm 45$ & $>$99\% & BNS & 1.4$\cdot10^{-5}$ & & \\ 
Observation Trigger$^{3}$ & 2019-04-25 10:09:02 & & & & & & 379 & 61 \\
Update Alert$^{4}$ & 2019-04-26 14:51:42 & 7461 (1378) & $146\pm 41$ & $>$99\% & BNS & 1.4$\cdot10^{-5}$ & & \\ 
GWTC-2.1$^{5}$ & & 8728 (2077) & $150^{+80}_{-60}$ & 78\% & BNS & 0.034 & & \\
\hline
\multicolumn{9}{c}{GW190426\_152155} \\
\hline
GW Detection$^{6}$ & 2019-04-26 15:21:55 & & & & & & & \\
Preliminary Alert$^{6}$ & 2019-04-26 15:47:06 & 1932 (472) & $423\pm 128$ & 86\% & BNS & 0.60 & & \\ 
Initial Alert$^{7}$ & 2019-04-26 15:53:09 & 1932 (472) & $423\pm 128$ & 86\% & BNS & 0.60 & & \\ 
Observation Trigger$^{3}$ & 2019-04-26 15:58:55 & & & & & & 20 & 6 \\
Update Alert$^{8}$ & 2019-04-27 11:52:26 & 1131 (214) & $377\pm 100$ & 86\% & BNS & 0.60 & & \\ 
Observation Trigger$^{3}$ & 2019-04-27 16:46:21 & & & & & & 355 & 48 \\
Update Alert$^{9}$ & 2019-05-06 16:21:43 & 1131 (214) & $377\pm 100$ & 86\% & NSBH & 0.60 & & \\ 
Update Alert$^{10}$ & 2019-08-29 14:46:43 & 1131 (214) & $377\pm 100$ & 42\% & BNS & 0.60 & & \\ 
GWTC-2$^{11}$ & & 1393 (221) & $370^{+180}_{-160}$ & $<50\%$ & NSBH & 1.4 & &\\
\hline
\multicolumn{9}{c}{S190510g} \\
\hline
GW Detection$^{12}$ & 2019-05-10 02:59:39 & & & & & & & \\
Initial Alert$^{13}$ & 2019-05-10 05:24:59 & 3462 (575) & $269\pm 108$ & 98\% & BNS & 0.026 & & \\ 
Update Alert$^{14}$ & 2019-05-10 10:22:54 & 1166 (31) & $227\pm 92$ & 98\% & BNS & 0.026 & & \\ 
Observation Trigger$^{3}$ & 2019-05-10 16:01:17 & & & & & & 50 & 17 \\
Update Alert$^{15}$ & 2019-05-10 20:43:51 & 1166 (31) & $227\pm 92$ & 85\% & BNS & 0.026 & & \\ 
Update Alert$^{16}$ & 2019-05-11 20:18:44 & 1166 (31) & $227\pm 92$ & 42\% & BNS & 0.28 & & \\ 
\hline
\multicolumn{9}{c}{GW190728\_064510} \\
\hline
GW Detection$^{17}$ & 2019-07-28 06:45:10 & & & & & & & \\
Preliminary Alert$^{17}$ & 2019-07-28 06:59:27 & 977 (211) & $786\pm 212$ & 100\% & BBH & 2.0$\cdot 10^{-4}$ & & \\
Preliminary Alert$^{17}$ & 2019-07-28 07:03:41 & 1169 (261) & $771\pm 209$ & $>$99\% & BBH & 2.0$\cdot 10^{-4}$ & & \\
Preliminary Alert$^{17}$ & 2019-07-28 07:39:04 & 1169 (261) & $771\pm 209$ & $>$99\% & Mass Gap & 7.7$\cdot 10^{-16}$ & & \\
Initial Alert$^{18}$ & 2019-07-28 07:50:42 & 543 (55) & $795\pm 197$ & $>$99\% & Mass Gap & 7.7$\cdot 10^{-16}$ & & \\ 
Observation Trigger$^{3}$ & 2019-07-28 09:35:19 & & & & & & 113 & 51 \\
Update Alert$^{19}$ & 2019-07-28 21:54:44 & 104 (24) & $874\pm 171$ & $>$99\% & BBH & 7.7$\cdot 10^{-16}$ & & \\ 
GWTC-2.1$^{20}$ & & 400 (56) & $880^{+260}_{-380}$ & 100\% & BBH & $<1.0\cdot 10^{-5}$ & & \\
\hline
\multicolumn{9}{c}{GW190814} \\
\hline
GW Detection$^{21}$ & 2019-08-14 21:10:39 & & & & & & & \\
Preliminary Alert$^{21}$ & 2019-08-14 21:31:40 & 772 (133) & $236\pm 53$ & 100\% & Mass Gap & 1.0$\cdot 10^{-17}$ & & \\ 
Initial Alert$^{22}$ & 2019-08-14 21:39:57 & 38 (7) & $276\pm 56$ & 100\% & Mass Gap & 1.0$\cdot 10^{-17}$ & & \\ 
Observation Trigger$^{3}$ & 2019-08-14 22:17:32 & & & & & & 256 & 60 \\
Update Alert$^{21}$ & 2019-08-14 23:56:23 & 38 (7) & $236\pm 53$ & 100\% & Mass Gap & 6.3$\cdot 10^{-26}$ & & \\ 
Observation Trigger$^{3}$ & 2019-08-15 06:26:49 & & & & & & 45 & 40 \\
Update Alert$^{23}$ & 2019-08-15 10:19:10 & 23 (5) & $267\pm 52$ & 100\% & NSBH & 6.3$\cdot 10^{-26}$ & & \\ 
Observation Trigger$^{3}$ & 2019-08-15 13:11:59 & & & & & & 349 & 66 \\
GWTC-2.1$^{24}$ & & 22 (4) & $230^{+40}_{-50}$ & 100\% & NSBH & $<1.0\cdot 10^{-5}$ & & \\
\hline
\multicolumn{9}{c}{S190822c} \\
\hline
GW Detection$^{25}$ & 2019-08-22 01:29:59 & & & & & & & \\
Preliminary Alert$^{25}$ & 2019-08-22 01:36:56 & 2767 (456) & $35\pm 10$ & $>$99\% & BNS & 1.9$\cdot 10^{-10}$ & & \\ 
Observation Trigger$^{3}$ & 2019-08-22 01:50:44 & & & & & & 38 & 36 \\
Retraction Alert$^{26}$ & 2019-08-22 03:12:19 & & & & & & & \\
\hline
\multicolumn{9}{c}{GW191216\_213338} \\
\hline
GW Detection$^{27}$ & 2019-12-16 21:33:38 & & & & & & & \\
Preliminary Alert$^{27}$ & 2019-12-16 21:50:10 & 300 (86) & $324\pm 78$ & 100\% & Mass Gap & 3.4$\cdot 10^{-16}$ & & \\ 
Initial Alert$^{28}$ & 2019-12-16 21:57:54 & 300 (86) & $324\pm 78$ & 100\% & Mass Gap & 3.4$\cdot 10^{-16}$ & & \\ 
Observation Trigger$^{3}$ & 2019-12-16 22:30:02 & & & & & & 73 & 42 \\
Update Alert$^{29}$ & 2019-12-19 17:55:16 & 253 (68) & $376\pm 70$ & $>$99\% & Mass Gap & 3.4$\cdot 10^{-16}$ & & \\
Update Alert$^{30}$ & 2019-12-23 18:01:58 & 253 (68) & $376\pm 70$ & $>$99\% & BBH & 3.4$\cdot 10^{-16}$ & & \\
GWTC-3$^{31}$ & & 480 (87) & $340^{+120}_{-130}$ & 99\% & BBH & $<1.0\cdot 10^{-5}$ & & \\
\hline
\hline
\multicolumn{9}{c}{Observation Run 4} \\
\hline
\hline
\multicolumn{9}{c}{S240422ed} \\
\hline
GW Detection$^{32}$ & 2024-04-22 21:35:13 & & & & & & & \\
Preliminary Alert$^{32}$ & 2024-04-22 21:36:09 & 376 (87) & $219\pm62$ & $>$99\% & NSBH & $9.8\cdot 10^{-6}$ & & \\
Preliminary Alert$^{32}$ & 2024-04-22 21:41:06 & 473 (133) & $205\pm61$ & $>$99\% & NSBH & $9.8\cdot 10^{-6}$ & & \\
Initial Alert$^{33}$ & 2024-04-22 22:21:02 & 404 (103) & $214\pm64$ & $>$99\% & NSBH & $9.8\cdot 10^{-6}$ & & \\
Observation Trigger$^{3}$ & 2024-04-22 22:23:48 & & & & & & 60 & 30 \\
Update Alert$^{34}$ & 2024-04-23 02:48:25 & 259 (72) & $188\pm43$ & $>$99\% & NSBH & $9.8\cdot 10^{-6}$ & & \\
Observation Trigger${^3}$ & 2024-04-23 05:07:44 & & & & & & 232 & 54 \\
\hline
\multicolumn{9}{c}{S250206dm} \\
\hline
GW Detection$^{35}$ & 2025-02-06 21:25:30 & & & & & & & \\
Preliminary Alert$^{35}$ & 2025-02-06 21:26:02 & 2139 (716) & $430\pm156$ & 27\% & BNS & 0.068 & & \\
Preliminary Alert$^{35}$ & 2025-02-06 21:30:52 & 1544 (486) & $409\pm139$ & 92\% & NSBH & 0.040 & & \\
Initial Alert$^{36}$ & 2025-02-06 21:49:19 & 1544 (486) & $409\pm139$ & 92\% & NSBH & 0.040 & & \\
Observation Trigger$^{3}$ & 2025-02-06 22:32:57 & & & & & & 71 & 36 \\
Update Alert$^{37}$ & 2025-02-06 23:39:39 & 1626 (459) & $359\pm125$ & 92\% & NSBH & 0.040 & & \\
Update Alert$^{37}$ & 2025-02-07 00:02:48 & 1626 (459) & $359\pm125$ & 92\% & NSBH & 0.040 & & \\
Update Alert$^{38}$ & 2025-02-07 02:46:08 & 910 (222) & $348\pm114$ & 92\% & NSBH & 0.040 & & \\
Update Alert$^{39}$ & 2025-02-08 11:06:27 & 547 (38) & $373\pm104$ & 92\% & NSBH & 0.040 & & \\
\enddata
\tablecomments{Event types with the highest probability per alert are presented. Number of galaxies observed for each observation trigger. $p_{astro}$ and FAR from the GWTC can vary, depending on the search pipeline. For consistency, we present the GstLAL pipeline \citep{gstlal_cannon_2021} estimates, as they are available for every event presented here. The O4b and O4c data has yet to be released to the GWTC.}
\tablenotetext{}{References: $^{1}$\url{https://gracedb.ligo.org/superevents/S190425z}, 
$^{2}$\cite{2019GCN.24168_initial_loc_s190425z}, 
$^{3}$Table \ref{tab:observations_semilist}, $^{4}$\cite{2019GCN.24228_updated_loc_s190425z}, 
$^{5}$\url{https://gwosc.org/eventapi/html/GWTC-2.1-confident/GW190425/v3/}, 
$^{6}$\url{https://gracedb.ligo.org/superevents/S190426c}, 
$^{7}$\cite{2019GCN.24237_initial__loc_s190426c},
$^{8}$\cite{2019GCN.24277_updated_loc_s190426c}, 
$^{9}$\cite{2019GCN.24411_updated_clas_s190426c}, 
$^{10}$\cite{2019GCN.25549_updated_clas_s190426c}, 
$^{11}$\url{https://gwosc.org/eventapi/html/GWTC-2/GW190426_152155/v1/}, 
$^{12}$\url{https://gracedb.ligo.org/superevents/S190510g}, 
$^{13}$\cite{2019GCN.24442_initial_loc_s190510g}, 
$^{14}$\cite{2019GCN.24448_updated_loc_s190510g}, 
$^{15}$\cite{2019GCN.24489_updated_sig_s190510g}, 
$^{16}$\cite{2019GCN.24462_updated_class_s190510g}, 
$^{17}$\url{https://gracedb.ligo.org/superevents/S190728q}, 
$^{18}$\cite{2019GCN.25187_initial_loc_s190728q}, 
$^{19}$\cite{2019GCN.25208_updated_loc_s190728q}, 
$^{20}$\url{https://gwosc.org/eventapi/html/GWTC-2.1-confident/GW190728_064510/v2/}, 
$^{21}$\url{https://gracedb.ligo.org/superevents/S190814bv}, 
$^{22}$\cite{2019GCN.25324_initial_loc_s190814bv}, 
$^{23}$\cite{2019GCN.25333_updated_loc_s190814bv}, 
$^{24}$\url{https://gwosc.org/eventapi/html/GWTC-2.1-confident/GW190814/v3/}, 
$^{25}$\url{https://gracedb.ligo.org/superevents/S190822c},  
$^{26}$\cite{2019GCN.25442_retract_s190822c}, 
$^{27}$\url{https://gracedb.ligo.org/superevents/S191216ap}, 
$^{28}$\cite{2019GCN.26454_initial_loc_s191216ap}, 
$^{29}$\cite{2019GCN.26505_updated_loc_s191216ap}, 
$^{30}$\cite{2019GCN.26570_updated_class_s191216ap}, 
$^{31}$\url{https://gwosc.org/eventapi/html/GWTC-3-confident/GW191216_213338/v1/},
$^{32}$\url{https://gracedb.ligo.org/superevents/S240422ed},
$^{33}$\cite{2024GCN.36236....1L},
$^{34}$\cite{2024GCN.36240....1L},
$^{35}$\url{https://gracedb.ligo.org/superevents/S250206dm},
$^{36}$\cite{2025GCN.39175....1L},
$^{37}$\cite{2025GCN.39178....1L},
$^{38}$\cite{2025GCN.39184....1L},
$^{39}$\cite{2025GCN.39231....1L}.}
\end{deluxetable*}

\begin{deluxetable*}{lcccccccccc}
\label{tab:observations_semilist}
\tablecaption{Las Cumbres observations obtained for the O3 and O4 events followed here.}
\tablehead{
\colhead{Event} &
\colhead{GLADE} &
\colhead{R.A.} &
\colhead{Dec.} &
\colhead{Luminosity} &
\colhead{$m_{B}$} &
\colhead{$L_{B}/L_{B}^{*}$} &
\colhead{Exposure Start} &
\colhead{Telescope} &
\colhead{Filter} &
\colhead{Limiting} \\
\colhead{ID} &
\colhead{ID} &
\colhead{} &
\colhead{} &
\colhead{Distance [Mpc]} &
\colhead{} &
\colhead{} &
\colhead{[UT]} &
\colhead{} &
\colhead{} &
\colhead{Mag. ($5\sigma$)}
}
\startdata
S190425z & 286940 & 252.337000 & -17.644520 & 102 & 11.60 & 12.63 & 2019-04-26 10:09:30 & OGG 2\,m & $g$ & 22.96 \\
S190425z & 286940 & 252.337000 & -17.644520 & 102 & 11.60 & 12.63 & 2019-04-25 19:03:31 & COJ 1\,m & $r$ & 19.67 \\
S190425z & 286940 & 252.337000 & -17.644520 & 102 & 11.60 & 12.63 & 2019-04-25 18:57:49 & COJ 1\,m & $i$ & 19.23 \\
S190425z & 286940 & 252.337000 & -17.644520 & 102 & 11.60 & 12.63 & 2019-04-25 19:09:13 & COJ 1\,m & $g$ & 20.16 \\
S190425z & 286940 & 252.337000 & -17.644520 & 102 & 11.60 & 12.63 & 2019-04-26 09:55:11 & OGG 2\,m & $i$ & 22.41 \\
S190425z & 286940 & 252.337000 & -17.644520 &  102 & 11.60 & 12.63 & 2019-04-26 10:03:47 & OGG 2\,m & $r$ & 23.09 \\
S190425z & 6752 & 253.723526 & -16.952021 & 123 & 13.35 & 3.63 & 2019-04-25 22:45:23 & CPT 1\,m & $i$ & 20.79 \\
\enddata
\tablecomments{$m_B$ denotes the apparent $B$-band magnitude of the target galaxy, taken from GLADE and \ensuremath{L_{B}^{*}} corresponds to \ensuremath{M^{*}_{B}=-20.7}, following \cite{Arcavi_2017_o2_strategy}.  Telescope names correspond to the codes in Table \ref{tab:lco_observatories}. Limiting magnitudes are calculated by {\tt{LCOGTSNpipe}} per image. This table is published in its entirety in the machine readable format. A portion is shown here for guidance regarding its form and content.}
\end{deluxetable*}

\section{Analysis}
\label{sec:analysis}
Despite the lack of optical counterparts discovered in O3 and O4, we wish to assess the effectiveness of our follow-up strategy and facility. First, we analyze the response time of Las Cumbres as the time from the event to the alert to triggering to start of observations (Section \ref{subsec:response_time_analysis}). Next, we evaluate the completeness of our galaxy targeted follow-up, in terms of the galaxy luminosity fraction covered out of the galaxies prioritized inside the localization volume that are brighter than the Schechter Function median value (Section \ref{subsec:galaxy_targeted_followup_analysis}). Finally, we examine the follow-up depth achieved by the triggered observations (Section \ref{subsec:followup_depth_analysis}).
\subsection{Response Time}
\label{subsec:response_time_analysis}
We define the overall ``Total Response Time'' as the time between the detection of a GW event to the execution of the first Las Cumbres observation. We subdivide this period into three intervals: (i) ``Time to Alert'', which is the time between the candidate merger and the issuing of the first alert with a non-zero probability for an EM signal (hereafter ``EM alert''), (ii) ``Time to Trigger'', which is the time between the first EM alert and the first submission of an observation request to the Las Cumbres scheduler\footnote{The decision to trigger follow-up for S190510g was made following the first update alert, since it reduced the 90\% (50\%) localization area by a factor of $\sim$3 ($\sim$20), so we use this alert instead of the initial alert for this case. Similarly, the first preliminary alert for S250206dm had $p_{astro}=27\%$, so we use the second preliminary alert for this case, which had a $p_{astro}$ of 92\%.}, and (iii) ``Time to Observations'', which is the time between the first observation request submission and the first observation on sky. The Las Cumbres response time is thus the sum of the Time to Trigger and the Time to Observations. This is thus the shortest time it took to start the search, but since the counterpart can be detected in the $n$-th galaxy after the start of the search, the estimated average time to the detection of a counterpart will be $n$ times the time spent observing each galaxy, divided by the average number of telescopes per site that can observe different galaxies simultaneously. Assuming three exposures of 300 seconds per galaxy and three telescopes per site that work in parallel, a detectable counterpart can be found in a few minutes to an hour in high priority galaxies ($n\leq10$), to a few hours for lower priority galaxies, after the start of observations.  We provide an observation timeline for each event, including periods of telescope unavailability, retrieved using the Las Cumbres Application Programming Interface (API)\footnote{\url{https://developers.lco.global/\#get-telescope-states}}.
\subsection{Galaxy Prioritization and Coverage}
\label{subsec:galaxy_targeted_followup_analysis}
We define galaxy luminosity coverage for each event as the fraction of $B$-band luminosity observed by us, $L_{B}$, out of the cumulative $B$-band luminosity of all prioritized galaxies that are brighter than the Schechter function median value (see Section \ref{sec:followup_strategy}) in the most recent alert preceding the final observation trigger, $L_{B,c}$. We distinguish between two types of coverage: (i) ``Total Luminosity Coverage'', where repeat observations of the same galaxies (either at different times or different wavelengths) are included, and (ii) ``Unique Luminosity Coverage'', where each galaxy is counted only once, regardless of the number of times it was observed.   The Total Luminosity Fraction is an upper limit for the achievable luminosity fraction if all observations were unique.

\subsection{Follow-up Depth}
\label{subsec:followup_depth_analysis}
We examine the 5$\sigma$ non-detection limits of each image obtained and compare them to GW170817 observations in the $g$, $r$, and $i$ bands taken from the \cite{Arcavi_early_hours} compilation of observations by \cite{andreoni_gw170817_obs, arcavi_gw170817_obs, coulter_gw170817_obs, cowperthwaite_gw170817_obs, diaz_gw170817_obs, drout_gw170817_obs, evans_gw170817_obs, hu_gw170817_obs, kasliwal_gw170817_obs, lipunov_gw170817_obs, pian_gw170817_obs, shapee_gw170817_obs, smartt_gw170817_obs, tanvir_gw170817_obs, troja_gw170817_obs, utsumi_gw170817_obs, dlt40_gw170817_valenti} and \cite{pozanenko_gw170817_obs}. We use the InfraRed Survey Archive (IRSA) dust extinction service  queries\footnote{\url{https://astroquery.readthedocs.io/en/latest/ipac/irsa/irsa_dust/irsa_dust.html}} \citep{Irsa2022-ab} to correct all non-detection limits for Milky Way dust extinction at the center of each image by using the \cite{irsa_schlafly2011} galactic dust extinction estimates. We then translate all limits to absolute magnitude using the mean distance of the pixel that overlaps with the center of the image in the most recent localization map (see Table \ref{tab:events_summary}). We translate the GW170817 observations to absolute magnitudes using the mean distance estimate provided by \cite{gw170817}. We define the limits that are deep enough to observe a GW170817-like kilonova as ``GW17 limits''. To check how many GW17 limits there are for each event, we linearly interpolate the GW170817 observations between epochs of the same filter. We do not extrapolate the GW170817 data to before its first detection (i.e. to earlier than 0.47 days). We refer to follow-up observations conducted here prior to 0.47 days from the event as ``Early Time Limits''. We compare the Las Cumbres capabilities to the observed properties of GW170817, rather than to kilonova models, given the wide diversity of model predictions and their dependence on uncertain physical assumptions.

\begin{figure}
    \centering
    \includegraphics[width=0.5\textwidth]{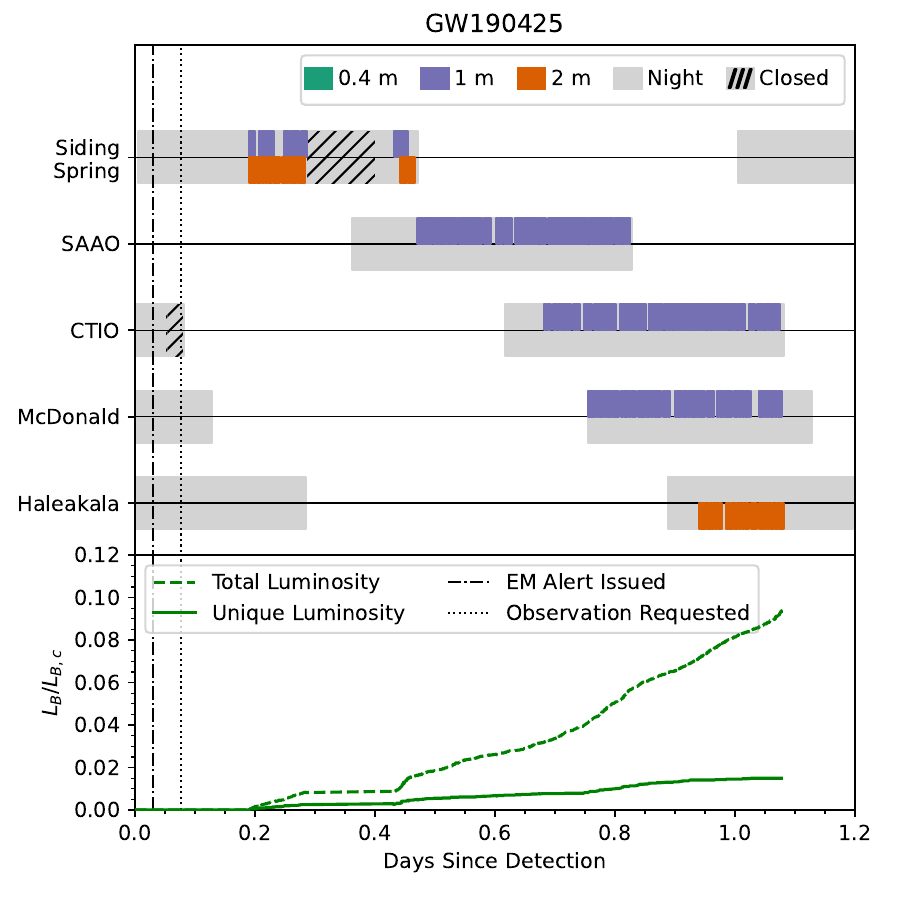}
    \caption{Top: Las Cumbres observation timeline for GW190425. On sky exposure blocks are shown in green, blue, and red for the 0.4\,m, 1\,m, and 2\,m telescopes, respectively. Grey regions are night periods, defined by astronomical twilight, and diagonal hash marks denote times when the telescopes at the site were in the ``not ok to open'' status (which could be due to weather or technical issues). Bottom: Luminosity fraction observed out of all prioritized galaxies.}
    \label{fig:s190425z_timeline}
\end{figure}
\begin{figure*}[t]
    \centering
    \includegraphics[width=0.5\textwidth]{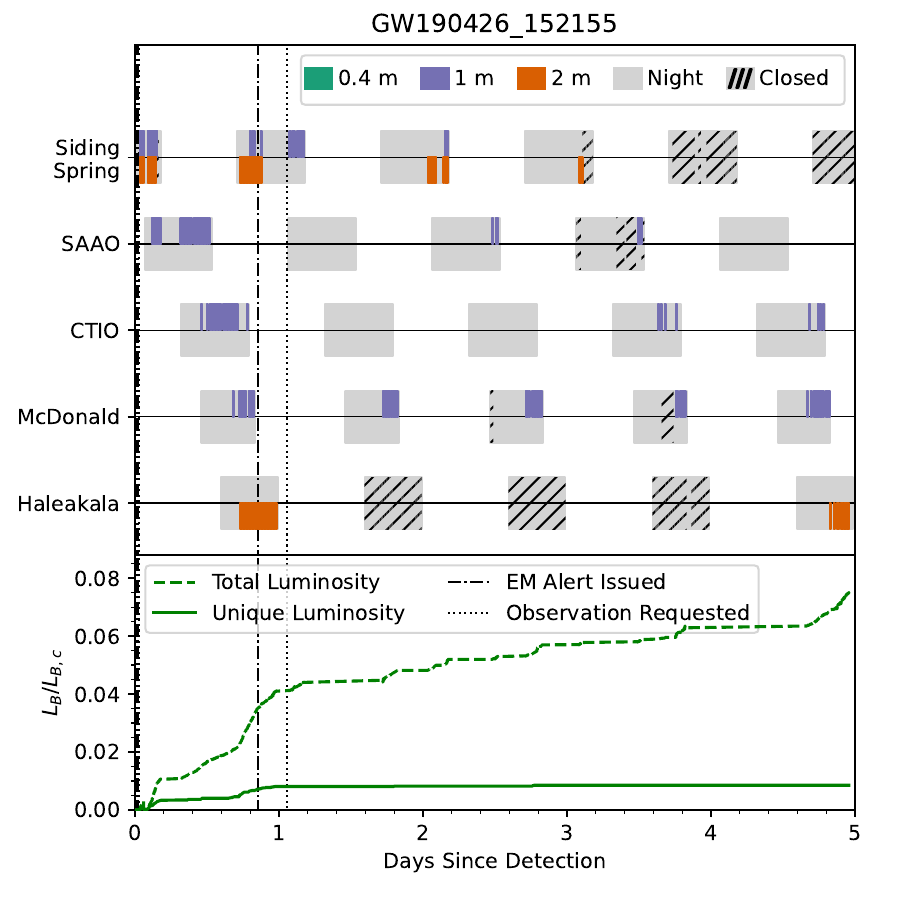}\,\,\includegraphics[width=0.5\textwidth]{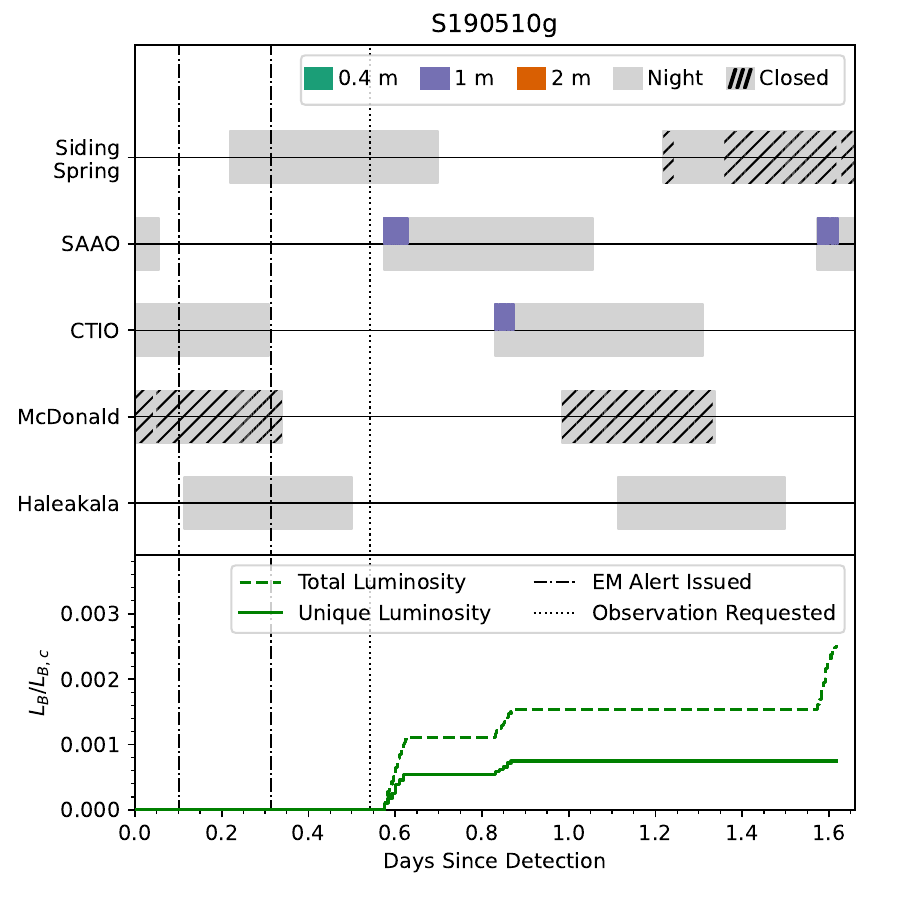}\\
    \includegraphics[width=0.5\textwidth]{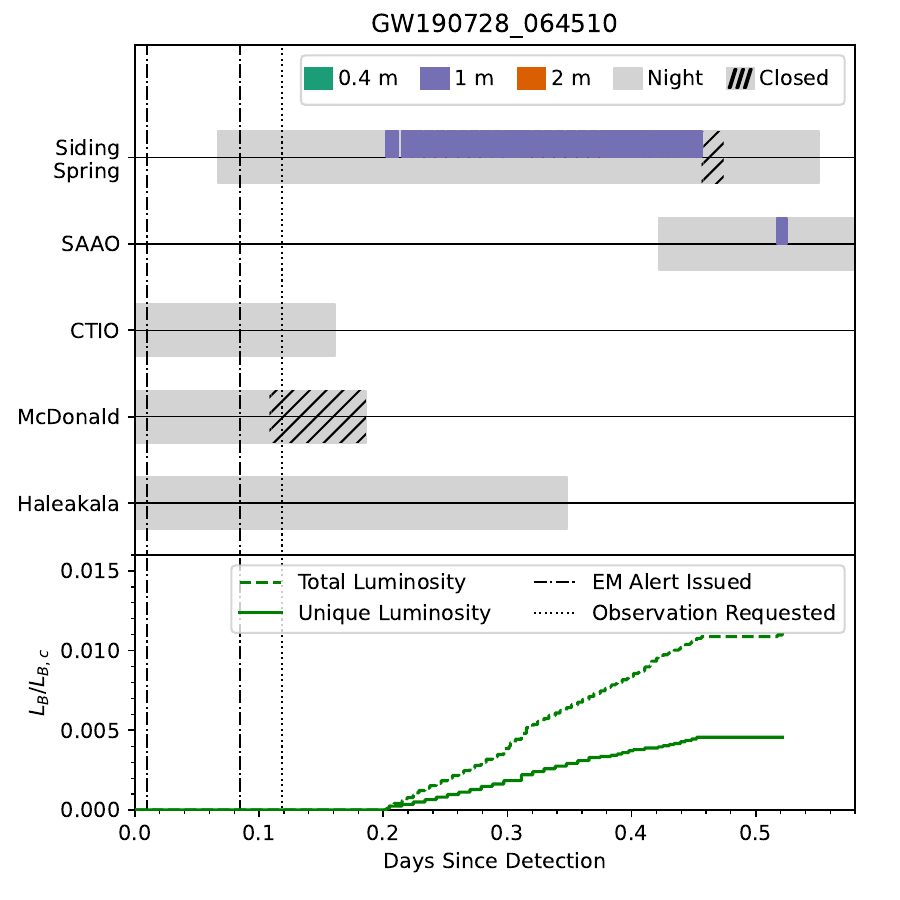}\,\,\includegraphics[width=0.5\textwidth]{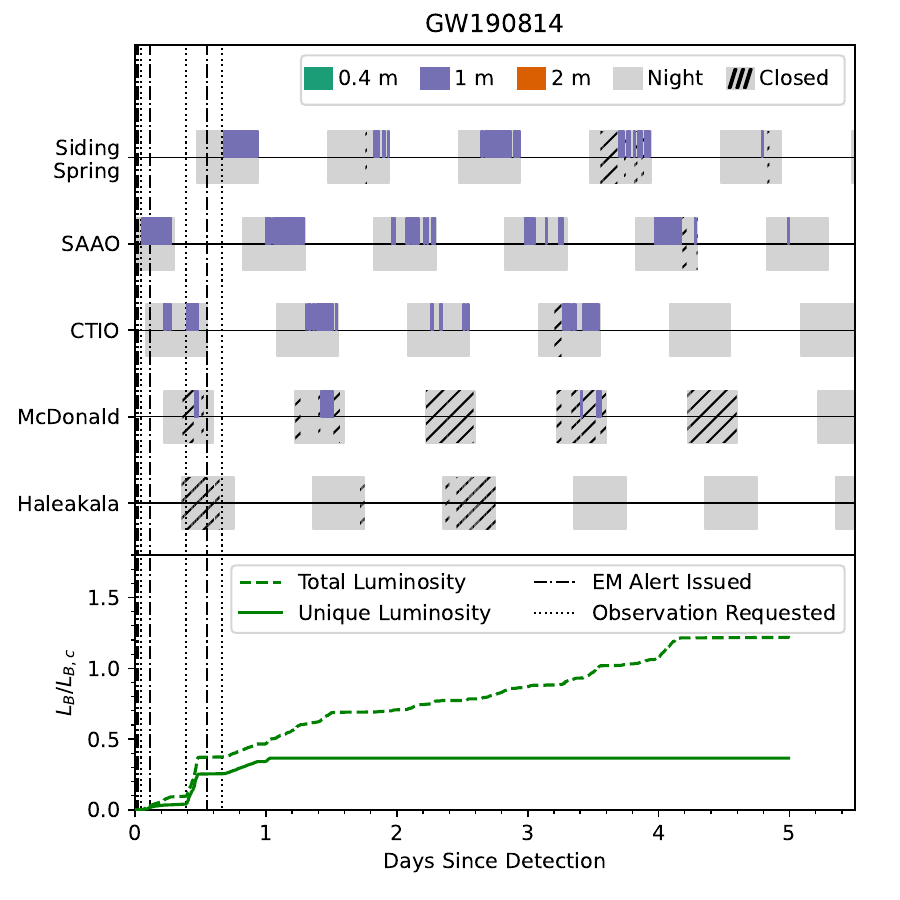}
    \caption{Same as Figure \ref{fig:s190425z_timeline}, but for GW190426\_152155 (top left), S190510g (top right), GW190728\_064510 (bottom left), and GW190814 (bottom right).}
    \label{fig:s190426c_timeline}
\end{figure*}

\begin{figure*}
    \centering
    \includegraphics[width=0.5\textwidth]{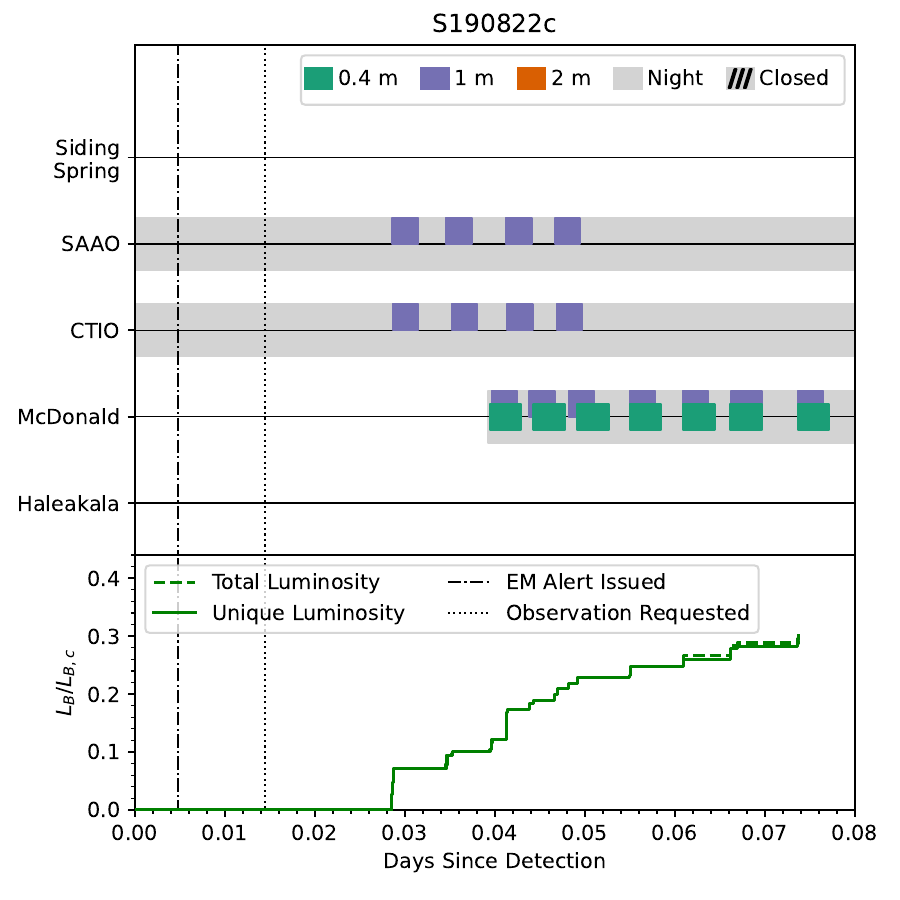}\,\,\includegraphics[width=0.5\textwidth]{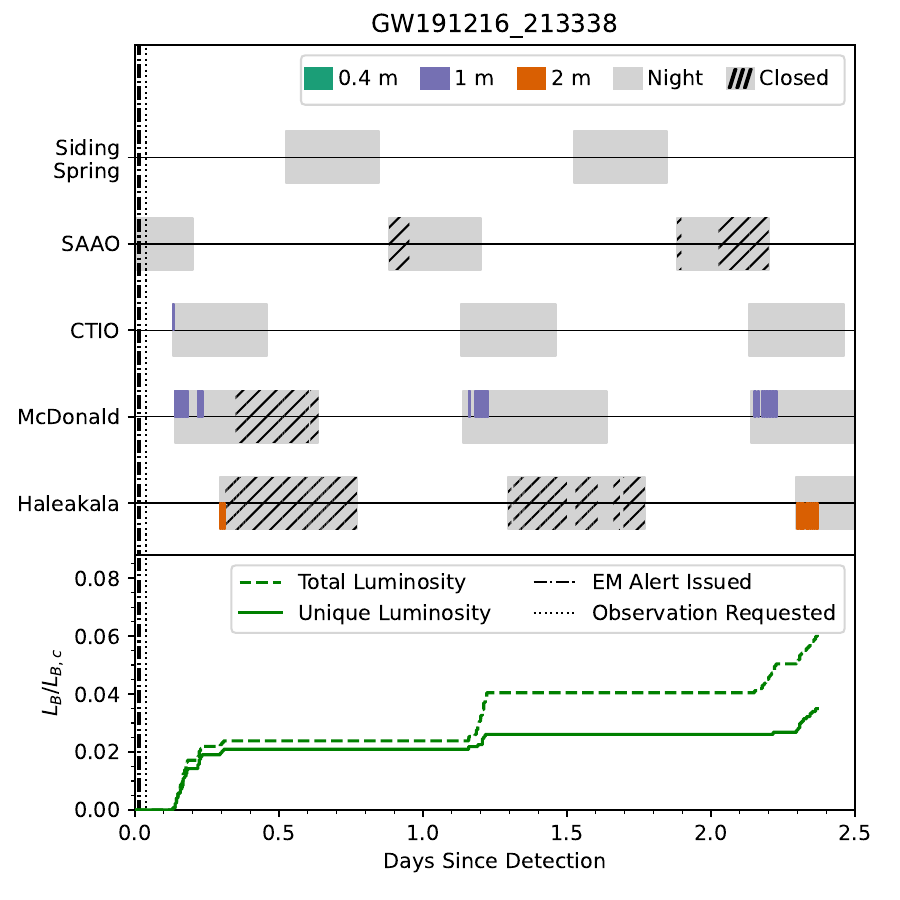}\\
    \includegraphics[width=0.5\textwidth]{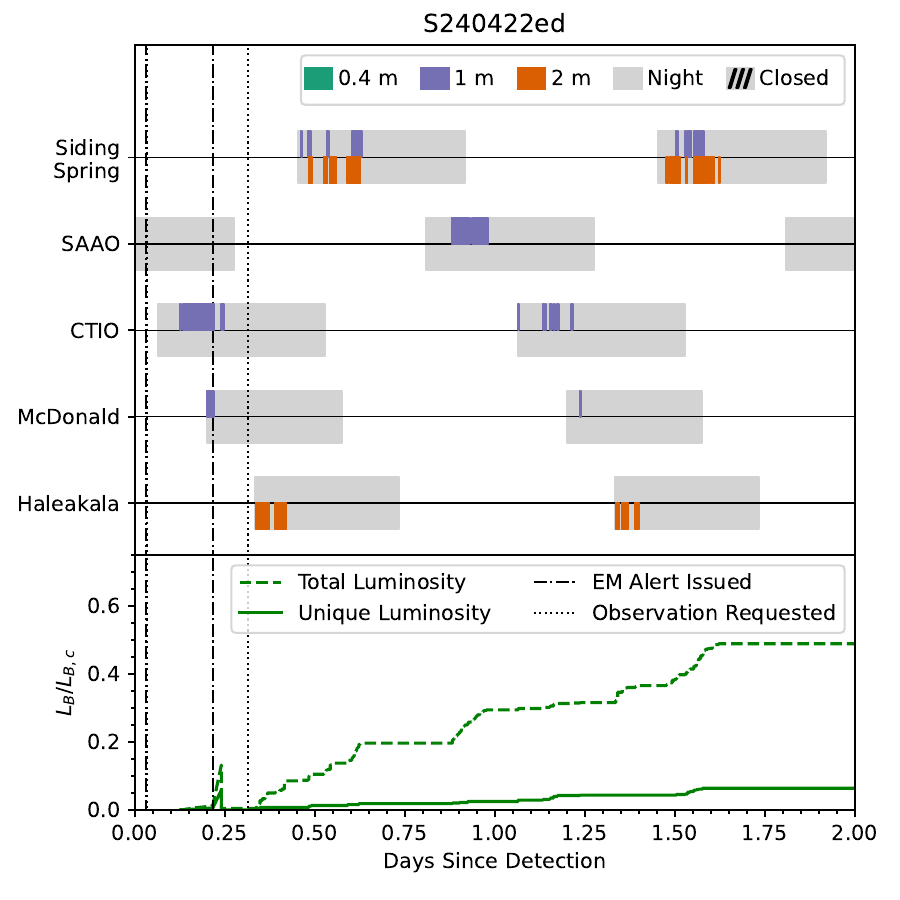}\,\,\includegraphics[width=0.5\textwidth]{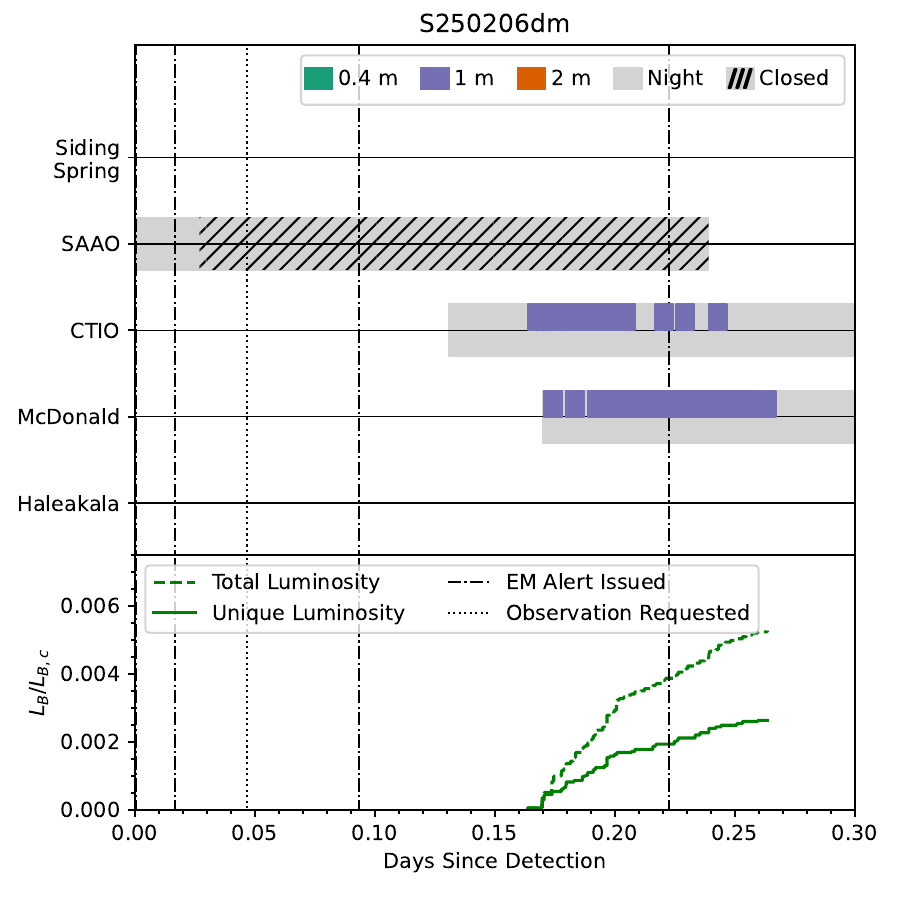}
    \caption{Same as Figure \ref{fig:s190425z_timeline}, but for S190822c (top left), GW191216\_213338 (top right), S240422ed (bottom left), S250206dm (bottom right). For S240422ed, we only show the first 2 days for clarity, but in practice the observations continued up to 8 days, contributing an additional fraction of $\sim$0.01 to the total luminosity.}
    \label{fig:s190728q_timeline}
\end{figure*}

\section{Results}
\label{sec:result_and_discussion}
The results of our analysis are summarized in Table \ref{tab:results_summary} and discussed below.

\begin{deluxetable*}{lccccccccc}[t]
\label{tab:results_summary}
\tablecaption{Summary of our results.}
\tablehead{\colhead{Event} & \colhead{Time to} & \colhead{Time to} & \colhead{Time to} & \colhead{Las Cumbres} & \colhead{Total Response} & \multicolumn{2}{c}{Luminosity Fraction} & \multicolumn{2}{c}{No. of Limits} \\
\colhead{} & \colhead{Alert} & \colhead{Trigger} & \colhead{Observations} & \colhead{Response Time} & \colhead{Time} & \colhead{Unique} & \colhead{Total} & \colhead{GW17} & \colhead{All} \\
\colhead{} & \colhead{[Hours]} & \colhead{[Hours]} & \colhead{[Hours]} & \colhead{[Hours]} & \colhead{[Hours]} & \colhead{} & \colhead{} & \colhead{} & \colhead{}}
\startdata
\hline
\hline
\multicolumn{9}{c}{O3} \\ 
\hline
\hline
GW190425 & 0.71 & 1.13 & 2.70 & 3.83 & 4.54 & 0.015 & 0.094 & 297 & 379 \\
GW190426\_152155 & 0.42 & 0.20 & 0.20 & 0.39 & 0.81 & 0.0086 & 0.0763 & 44 & 375 \\
S190510g & 7.39 & 5.64 & 0.76 & 6.40 & 13.79 & 0.00075 & 0.00253 & 43 & 50 \\
GW190728\_064510 & 0.90 & 1.94 & 2.02 & 3.96 & 4.86 & 0.0046 & 0.0111 & 0 & 113 \\
GW190814 & 0.35 & 0.76 & 0.21 & 0.97 & 1.32 & 0.37 & 1.24 & 40 & 650 \\
S190822c & 0.12 & 0.23 & 0.34 & 0.56 & 0.68 & 0.29 & 0.31 & 0 & 38 \\
GW191216\_213338 & 0.28 & 0.66 & 2.21 & 2.87 & 3.15 & 0.035 & 0.061 & 2 & 73 \\\hline
O3 Median & 0.42 & 0.76 & 0.76 & 2.87 & 3.15 & & \\
\hline
\hline
\multicolumn{9}{c}{O4} \\ 
\hline
\hline
S240422ed & 0.016 & 0.80 & 2.20 & 3.00 & 3.01 & 0.065 & 0.51 & 22 & 292 \\
S250206dm & 0.090 & 1.03 & 2.80 & 3.84 & 3.93 & 0.0026& 0.0053 & 0 & 71 \\\hline
O4 Median & 0.053 & 0.91 & 2.50 & 3.42 & 3.47 & & \\
\hline
\hline
O3+O4 Median & 0.35 & 0.80 & 2.02 & 3.00 & 3.15 & & \\
\enddata
\tablecomments{See Section \ref{sec:analysis} for information regarding each column.}
\end{deluxetable*}

\subsection{Response Time}
\label{subsec:response_time}
\begin{figure}
\hspace{-0.03\textwidth}
\includegraphics[width=0.5\textwidth]{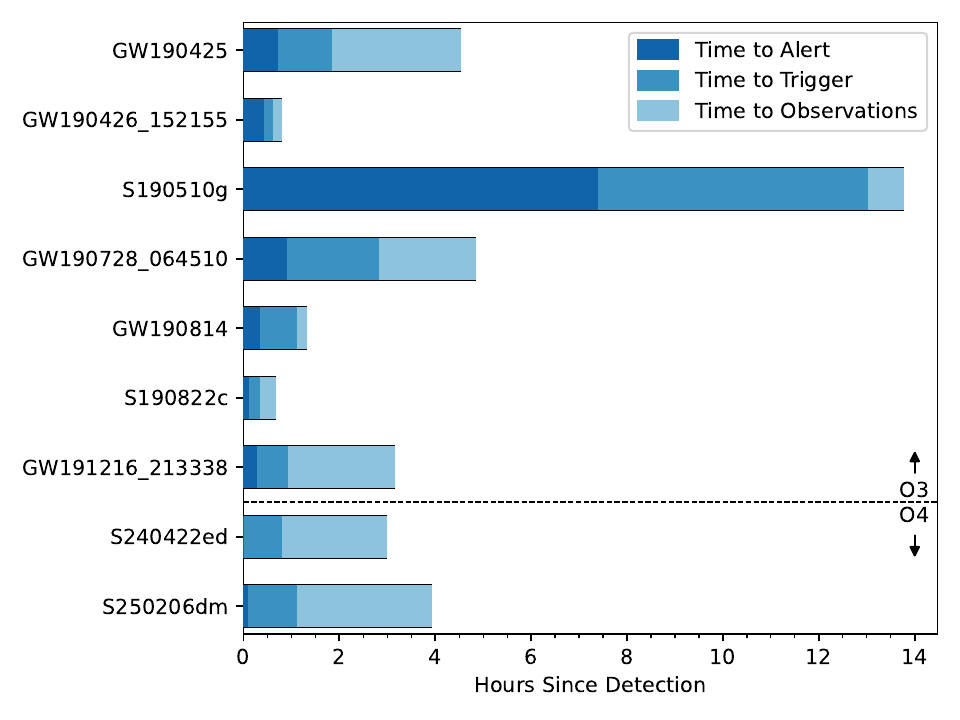}
\caption{Total response time and its constituents for each event. The times to alert shown correspond to the first EM alert issued (either preliminary or initial), except for S190510g  and S250206dm, where the time is from the first update and second preliminary EM alerts, respectively (see text for details).}
\label{fig:response_time_summary}
\end{figure}

\begin{figure}
\includegraphics[width=0.5\textwidth]{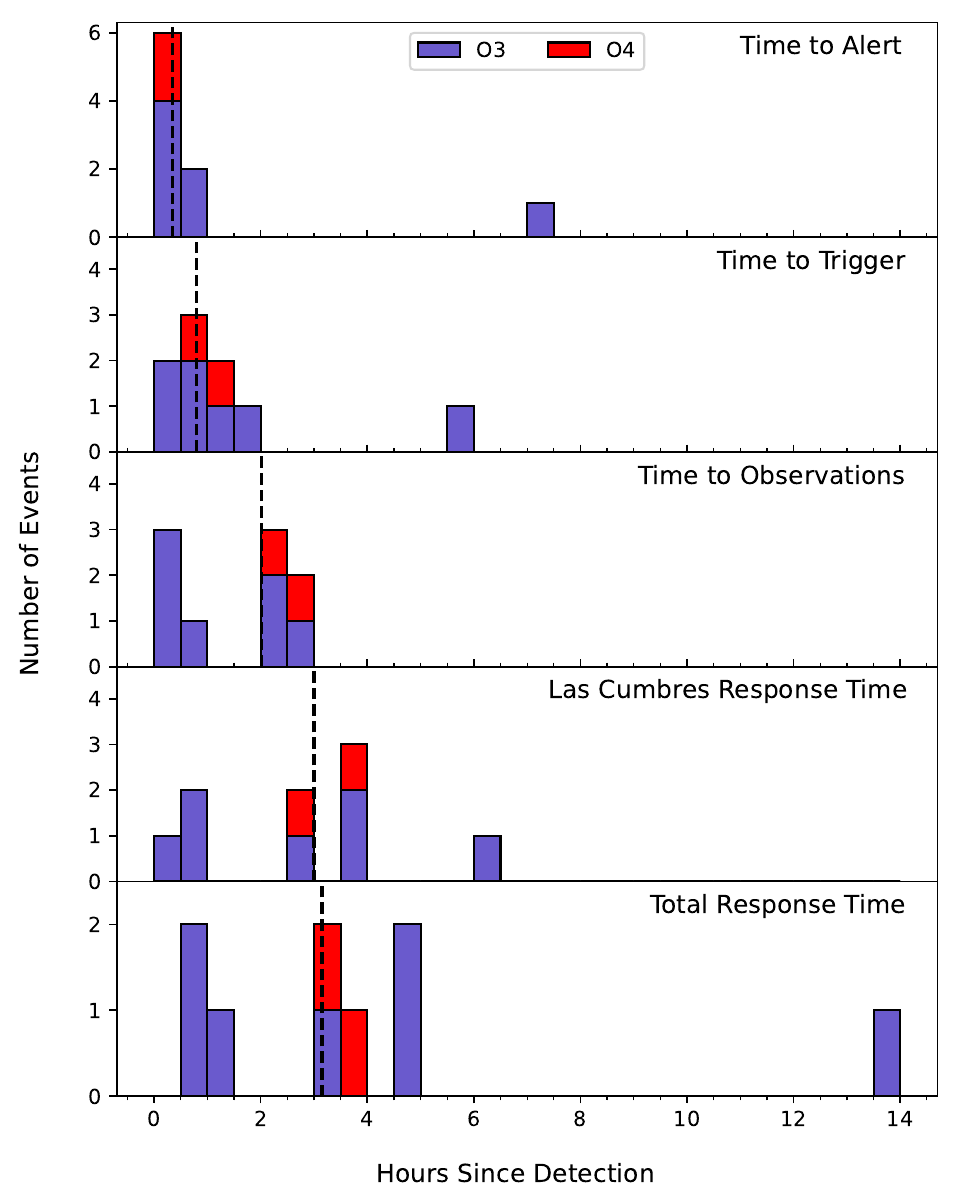}
\caption{From top to bottom: Time to Alert, to Trigger, and to Observations, Las Cumbres response time and Total Response Time histograms. Values for O3 and O4 events are presented in blue and red, respectively. Black dashed lines mark the O3+O4 median value for each panel.}
\label{fig:response_time_histograms}
\end{figure}

The top panels of Figures \ref{fig:s190425z_timeline}--\ref{fig:s190728q_timeline} show the observation timelines for the events followed here. Figures \ref{fig:response_time_summary} and \ref{fig:response_time_histograms} summarize the response time for all nine events. The median total response time is $3.15$ hours, with the quickest (0.68 hours) and slowest (13.79 hours) responses occurring for S190822c and S190510g, respectively (Table \ref{tab:results_summary}). Examining each interval separately, the medians for the time to alert, trigger, and observations are $0.35$ hours, $0.80$ hours, and $2.02$ hours, respectively. The response time is affected also by the GW alerts, which can change throughout an active event and affect our triggering decision. For example, the event GW190728\_064510 initially reported as a BBH merger with 100\% probability, and approximately 1 hour after detection was changed to a mass gap object with 50\% probability. The median of the Las Cumbres response time is $3.00$ hours. The time to observations can be affected by telescope availability due to weather, and source sky location.

In general, GW follow-up observation requests are submitted as disruptive ToO's, however the GW190425 follow-up was requested as a ``normal'' trigger, resulting in the first observation being obtained roughly $2.7$ hours after the request was submitted. GW190426\_152155, GW190814, and S190822c exhibit the fastest disruptive ToO response, with between 12 to 20 minutes of time to observations. The time to observations for GW191216\_213338 was $2.21$ hours, due to the localization being mostly in the northern hemisphere, while the only night-time site at the time of the alert, SAAO, is in the southern hemisphere. As soon as the night began at the CTIO site, first observations were taken (Fig. \ref{fig:s190728q_timeline}).

The decision on when to stop follow-up observations varied between events. By default, follow-up continued for up to 5 days after the GW detection, as demonstrated for GW190426\_152155 and GW190814 (Figs. \ref{fig:s190426c_timeline} and \ref{fig:s190728q_timeline}, respectively). The follow-up of GW190425 was stopped after $\sim$1.1 days (Fig. \ref{fig:s190425z_timeline}) due to the decision to prioritize follow-up of GW190426\_152155, as its localization area was smaller and more manageable. For S190510g, follow-up was stopped $\sim$1.6 days after the GW detection (Fig. \ref{fig:s190426c_timeline}) due to a decrease in $p_{astro}$ in subsequent alerts. In the cases of GW190728\_064510 and GW191216\_213338, the final update alerts reclassified the sources from Mass Gap objects to BBH mergers, prompting the end of follow-up after $\sim$0.55 and $\sim$2.4 days, respectively (Fig. \ref{fig:s190728q_timeline}). The follow-up of S190822c was terminated after a retraction alert was issued, approximately 2 hours post-detection (Fig. \ref{fig:s190728q_timeline})

\subsection{Luminosity Fraction Observed}
\label{subsec:luminosity_fraction_observed}
The bottom panels of Figures \ref{fig:s190425z_timeline}--\ref{fig:s190728q_timeline} show the total and unique cumulative luminosity fractions of the observations of each event. The best covered event is GW190814, achieving a unique luminosity coverage of 37\% within $\sim1.1$ days. Notably, GW190814 also had the smallest 90\% localization area, covering just 22 \sqd. 

In contrast, the least covered event is S190510g, which attained a unique (total) luminosity coverage of 0.075\% (0.25\%) after $\sim1.5$ days. This event also involved the highest number of prioritized galaxies, exceeding 25,000. The remaining events all exhibit similarly low total luminosity coverage (except for S190822c, though it was retracted after less than two hours).

The galaxy-targeted strategy proposed by \cite{Gehrels_2016_galaxy_strategy} assumed nearby and/or well-localized sources (search volumes of 500 \sqd\ \ensuremath{\times} 60 Mpc, 100 \sqd\ \ensuremath{\times} 120 Mpc, and 20 \sqd\ \ensuremath{\times} 180 Mpc for progressively sensitive runs of the GW detectors). This is consistent with the GW170817 localization, even without considering GRB170817A, and with the GW190814 localization. However, as shown in Table \ref{tab:events_summary}, most events followed here surpass these assumptions in either distance, area, or both, giving larger volumes by one or two orders of magnitudes. This results in $10^{3}-10^{5}$ galaxies being prioritized. Consequently, the observed luminosity fractions remained below a few percent in most cases. 

The difference between the total and unique luminosity fraction covered is usually caused by galaxies being observed twice or thrice in multiple filters, which can add 5-10 minutes of observation for a single target. For example, S190822c was observed solely in the $g$ filter, leading to a total and unique luminosity fraction of similar values (31\% and 29\%, respectively).

\subsection{Follow-up Depth}
\label{subsec:followup_depth}

\begin{figure} 
\includegraphics[width=0.5\textwidth]{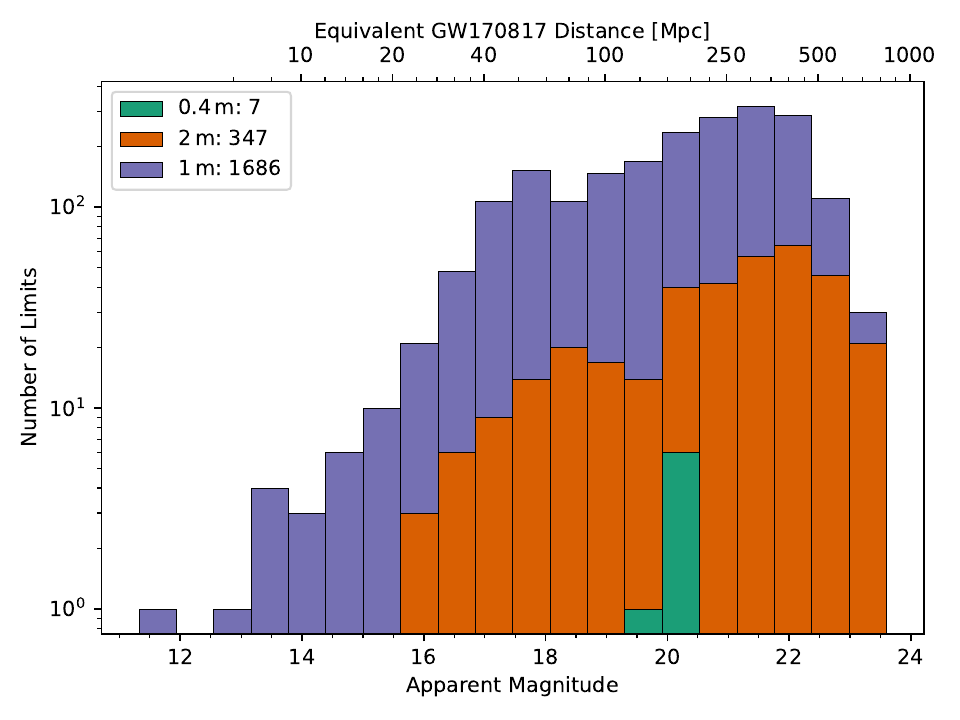}
    \caption{Histograms of apparent magnitude 5$\sigma$ limits achieved by the 1\,m, 2\,m, and 0.4\,m telescope classes for the observations presented here. The top axis indicates the equivalent distance at which a GW170817-like kilonova (having a $g$-band peak absolute magnitude of $-16$) would be detected at the corresponding apparent magnitude. The legend lists the total number of images acquired by each telescope class.}
    \label{fig:apparent_magnitudes_summary}
\end{figure}

Figure \ref{fig:apparent_magnitudes_summary} shows the distribution of apparent magnitude 5$\sigma$ limits achieved by different telescope classes.
More than half of the observations reached a depth of at least 21 magnitudes, sufficient to detect a GW170817-like event at peak brightness out to a distance of $\sim$250 Mpc.

\begin{figure*}[h] 
\includegraphics[width=\textwidth]{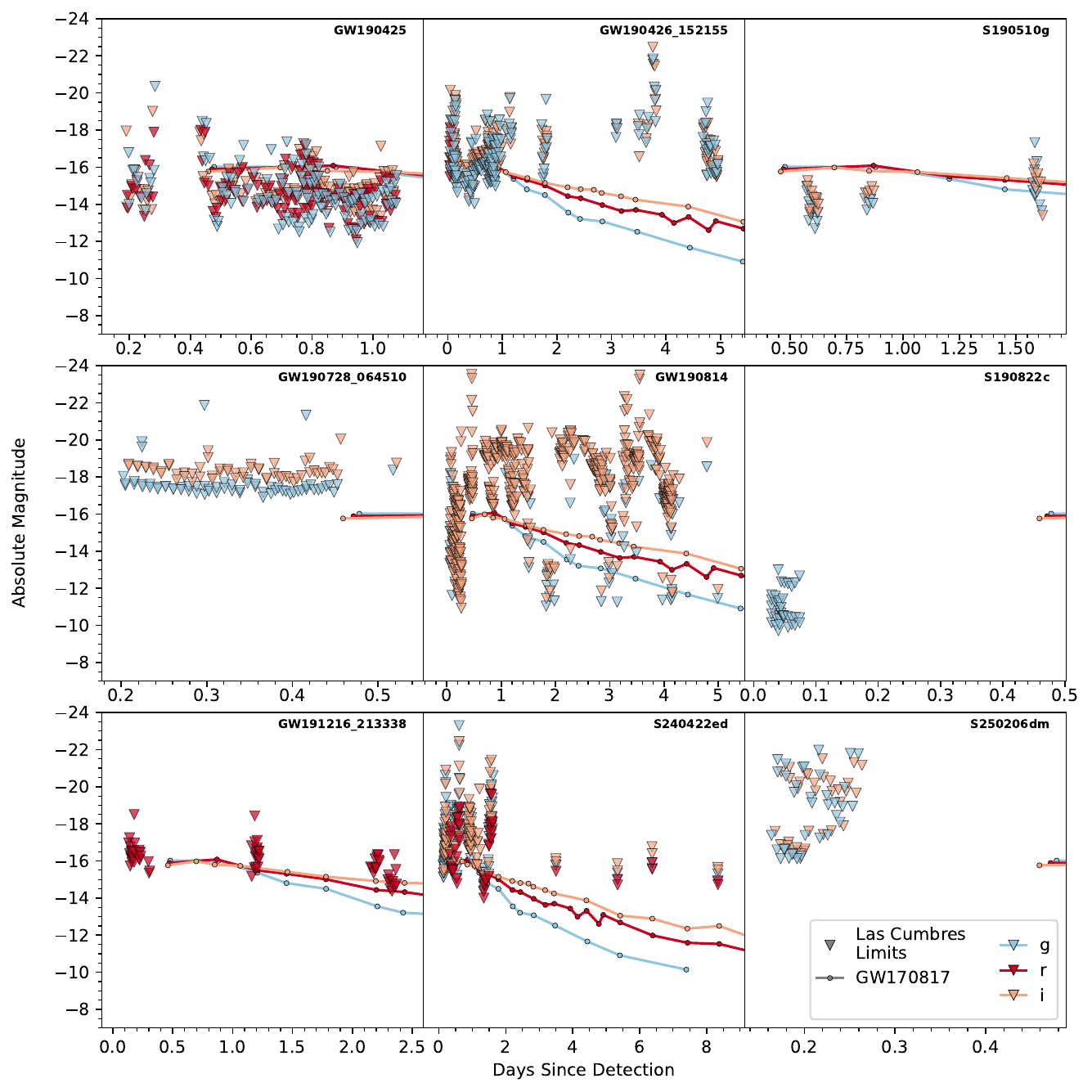}
\caption{Las Cumbres $5\sigma$ non-detection limits (triangles) for the observations presented here, and GW170817 observations (circles; see text for data references). We calculate the absolute magnitude of each limit using the latest distance estimate at that coordinate, provided for each event (see Section \ref{subsec:followup_depth}). All limits consider Milky Way dust extinction.}
\label{fig:o3_summary_magnitudes}
\end{figure*}

\begin{figure*}[!htbp] 
\includegraphics[width=\textwidth]{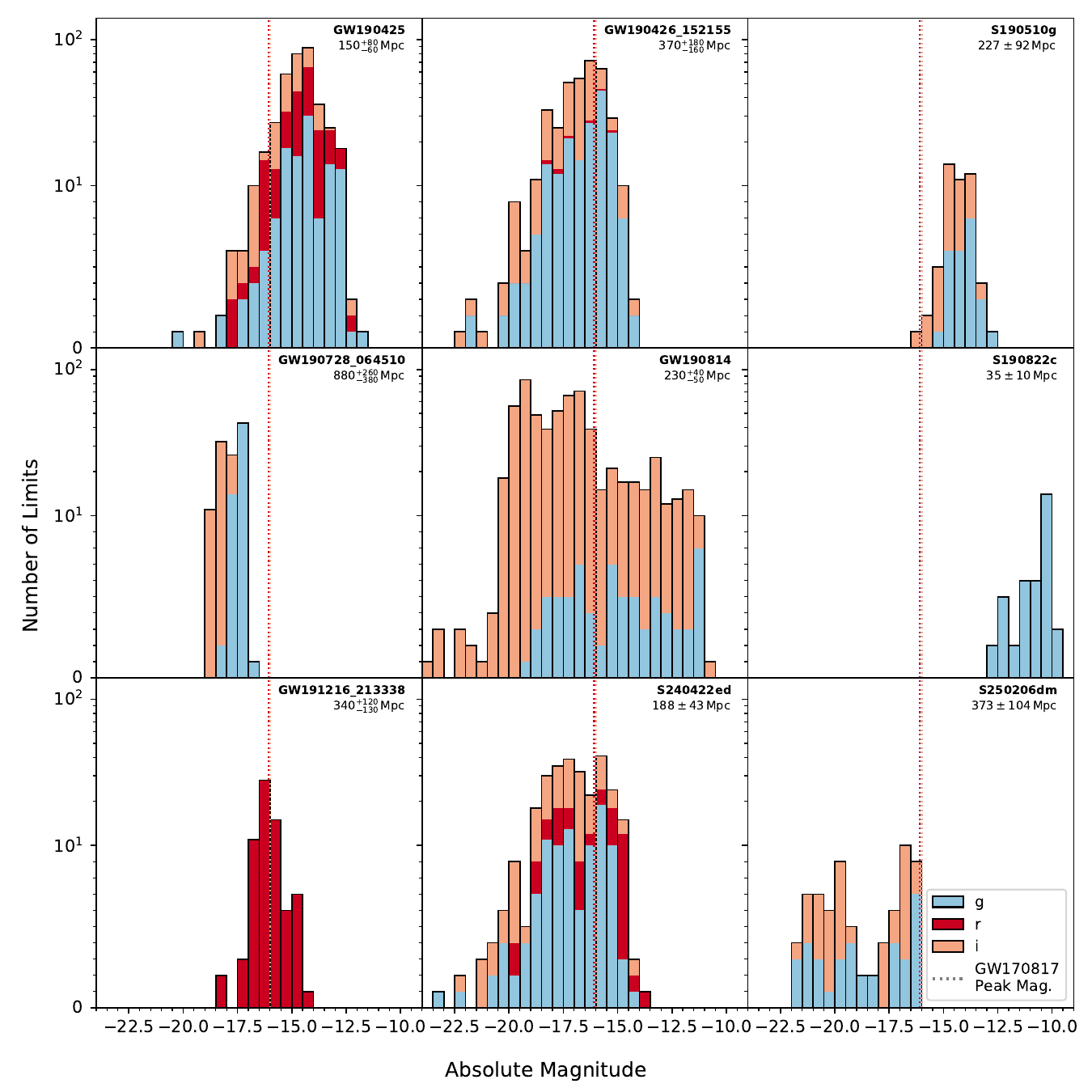}
\caption{Stacked histograms of the Las Cumbres $5\sigma$ non‑detection limits shown in Figure \ref{fig:o3_summary_magnitudes}, in the $g$, $r$ and $i$ bands denoted in green, red, and orange bars, respectively. The mean GW distances are presented on the top right corner of each panel, for reference. The y‑axis uses a hybrid scale: it is linear up to a value of 10 and logarithmic beyond that. Vertical dotted lines indicate the most luminous measured magnitude of the GW170817 kilonova in each filter, with line colors matching the respective bands.}
\label{fig:o3_summary_magnitudes_histograms}
\end{figure*}

Figures \ref{fig:o3_summary_magnitudes} and \ref{fig:o3_summary_magnitudes_histograms} present the absolute magnitude limits as a function of time with the GW170817 $g$, $r$, and $i$ band lightcurves for reference. The majority of the events had GW17 limits (i.e. limits deep enough to detect a GW170817-like kilonova at the appropriate distance), with GW190425 and S190510g having the highest fraction of such limits: 91\% and 86\% respectively. In contrast, the GW190728\_064510 and S190822c observations resulted in no GW17 limits. This discrepancy is due to the former being dominated by early-time limits and the latter exclusively featuring early-time limits (i.e. limits at times before the first detection of the GW170817 kilonova, during which we do not know how bright it was). Apart from S190510g, all other events included some early-time limits.

This does not mean that an EM counterpart to these events would have necessarily been observable, as EM counterparts to BNS and NSBH mergers are expected to vary in luminosity depending on their physical parameters and viewing angle. GW190728\_064510, S190822c, and GW191216\_213338 were not expected to exhibit an EM counterpart at all, either due to (eventually) being more likely BBH mergers or not being of astrophysical origin at all.

\section{Summary and Discussion}
\label{sec:summary}
We presented and analyzed Las Cumbres Observatory follow-up of seven GW events during O3 and O4. Given that no electromagnetic counterparts were expected for any of the events considered here \citep[with the exception of GW190425 which is discussed in detail in][]{keinan2025}, followup observations can not place meaningful constraints on the physical properties of potential counterparts. Therefore, this work focuses on evaluating the efficiency and response characteristics of the follow-up strategy itself. 

We find that the median response time from EM alert to first observation taken was 3.00 hours, with the shortest being 23 minutes. As shown by \cite{Arcavi_early_hours}, short response times are crucial, as the identification of the GW170817 kilonova $\sim$11 hours after the merger missed the early emission which could have distinguished between different kilonova models. Additionally, for the majority of cases, observations were deep enough to detect a GW170817-like kilonova. The ability of Las Cumbres to continuously observe across multiple sites, while compensating for local weather and equipment malfunctions, aligns well with the demands of searching for and following rapidly-evolving and short-lived kilonovae.

However, this analysis also highlights the limitations of a galaxy-targeted follow-up strategy when faced with large localization areas and distant sources. Only one event, GW190814, demonstrated a potential for complete galaxy coverage. This event was also the most similar to GW170817 in terms of localization area and number of prioritized galaxies. By contrast, the majority of events exhibited much larger localization volumes which included $10^{3}-10^{5}$ prioritized galaxies. The median 50\% and 90\% localization areas during O3 (O4) were 130 (242) \sqd\ and 740 (1021) \sqd, respectively, with a median distance of approximately 1115 (2180) Mpc. For events within 500 Mpc (which are more relevant for EM follow-up), the median 50\% and 90\% localization areas during O3 (O4) were 378 (30) \sqd\ and 3403 (113) \sqd, respectively\footnote{Here we consider all event types.}. As more detectors are added, localizations could get smaller, making galaxy-targeted strategies more feasible (as originally envisioned by \citealt{Gehrels_2016_galaxy_strategy}). However, this will still require sufficiently complete galaxy catalogs out to large distances. 

The interplay between future GW detector sensitivity, localization accuracy, and galaxy-catalog completeness will be critical in optimizing counterpart discovery and follow-up strategies. Currently, large-field tiling strategies can cover the larger GW localizations more efficiently, while dynamically-scheduled global facilities like Las Cumbres can shorten follow-up delay times. However, the lack of Las Cumbres reference images in real time could hinder identification of faint sources on top of bright backgrounds. While post-event reference images could allow for early-epoch recovery of known transients, this does limit the ability of a facility like Las Cumbres to provide the early discovery itself. Some overlap in search efforts, tailored to the nature of the real time data products of each survey, is therefore likely required and should be considered. This study thus emphasizes the need for collaboration between facilities with different capabilities \citep[as enabled, for example, by the Treasure Map;][]{wyatt2020_treasuremap} for obtaining rapid, deep and wide GW follow-up.

~\\
I.K. and I.A. acknowledge support from the Israel Science Foundation (grant No. 2752/19) and from the United States-Israel Binational Science Foundation (BSF; grant No. 2018166). I.A. further acknowledges support from the European Research Council (ERC) under the European Union’s Horizon 2020 research and innovation program (grant agreement No. 852097), and from the Pazy foundation (grant No. 216312). 
The Las Cumbres Observatory team is supported by National Science Foundation (NSF) grants AST-2308113 and AST-1911151.
A.H.N. acknowledges support from the NSF grant PHY-2309240.
D.P.'s research was funded in part by the Koret Foundation, the Kavli Institute for Particle Astrophysics and Cosmology at Stanford University, and by grant NSF PHY-2309135 to the Kavli Institute for Theoretical Physics (KITP). D.P. acknowledges further support from Israel Science Foundation (ISF) grant 541/17, and by grant 2018017 from the United States-Israel Binational Science Foundation (BSF).
D.J.S. is supported by NSF grants 2308181, 2407566, and 2432036.

This work makes use of observations from the Las Cumbres Observatory network, of data and software obtained from the Gravitational Wave Open Science Center (gwosc.org), a service of the LIGO Scientific Collaboration, the Virgo Collaboration, and KAGRA, version 2.3 of the GLADE galaxy catalog \citep{glade_catalog}, and the NASA/IPAC Infrared Science Archive, which is funded by the National Aeronautics and Space Administration and operated by the California Institute of Technology.

\software{{\qcr astropy} \citep{astropy2013, astropy2018, astropy2022}, {\qcr healpy} \citep{healpix}, {\qcr LCOGTSNpipe} \citep{valenti2016_lcogtsnpipe}, ligo.skymap, {\qcr matplotlib} \citep{matplotlib}, {\qcr numpy} \citep{numpy}, {\qcr pandas}, {\qcr scipy} \citep{scipy_virtanen2020}.}

\newpage
\bibliography{references}
\bibliographystyle{aasjournal}

\end{document}